\documentclass[sigconf]{acmart}

\AtBeginDocument{%
  \providecommand\BibTeX{{%
    \normalfont B\kern-0.5em{\scshape i\kern-0.25em b}\kern-0.8em\TeX}}}

\copyrightyear{2024}
\acmYear{2024}
\setcopyright{rightsretained}
\acmConference[CHI '24]{Proceedings of the CHI Conference on Human Factors in Computing Systems}{May 11--16, 2024}{Honolulu, HI, USA} \acmBooktitle{Proceedings of the CHI Conference on Human Factors in Computing Systems (CHI '24), May 11--16, 2024, Honolulu, HI, USA}
\acmDOI{10.1145/3613904.3642628}
\acmISBN{979-8-4007-0330-0/24/05}

\acmSubmissionID{8106}

\usepackage{xspace}
\usepackage{xcolor}
\usepackage{cleveref}

\definecolor{RoyalBlue}{HTML}{0071BC}

\newcommand{\ie}{{i.e.,}\xspace}
\newcommand{\eg}{{e.g.,}\xspace}

\newcommand{\system}{\textsc{Talaria}\xspace}
\newcommand{\tableview}{Table View\xspace}
\newcommand{\graphview}{Graph View\xspace}
\newcommand{\diffview}{Diff View\xspace}
\newcommand{\user}{Moira\xspace}

\newcommand{\location}{{Apple}\xspace}

\newcommand{\NumFormative}{12\xspace}
\newcommand{\NumSurveyParticipants}{26\xspace}
\newcommand{\NumInterviewParticipants}{7\xspace}

\begin{document}

\title[\textsc{\system{}}]{\textsc{\system{}}: Interactively Optimizing Machine Learning Models for Efficient Inference}

\settopmatter{authorsperrow=4}

\author{Fred Hohman}
\orcid{1234-5678-9012}
\affiliation{%
  \institution{Apple}
  \city{Seattle}
  \state{WA}
  \country{USA}
}
\email{fredhohman@apple.com}

\author{Chaoqun Wang}
\orcid{1234-5678-9012}
\affiliation{%
  \institution{Apple}
  \city{Beijing}
  \country{China}
}
\email{chaoqun\_wang@apple.com}

\author{Jinmook Lee}
\orcid{1234-5678-9012}
\affiliation{%
  \institution{Apple}
  \city{Cupertino}
  \state{CA}
  \country{USA}
}
\email{jinmook\_lee@apple.com}

\author{Jochen Görtler}
\authornote{Work done at Apple.}
\orcid{1234-5678-9012}
\affiliation{%
  \institution{Independent Researcher}
  \city{Walldorf}
  \country{Germany}
}
\email{me@jgoertler.com}

\author{Dominik Moritz}
\orcid{1234-5678-9012}
\affiliation{%
  \institution{Apple}
  \city{Pittsburgh}
  \state{PA}
  \country{USA}
}
\email{domoritz@apple.com}

\author{Jeffrey P. Bigham}
\orcid{1234-5678-9012}
\affiliation{%
  \institution{Apple}
  \city{Pittsburgh}
  \state{PA}
  \country{USA}
}
\email{jbigham@apple.com}

\author{Zhile Ren}
\orcid{1234-5678-9012}
\affiliation{%
  \institution{Apple}
  \city{Seattle}
  \state{WA}
  \country{USA}
}
\email{zhile\_ren@apple.com}

\author{Cecile Foret}
\orcid{1234-5678-9012}
\affiliation{%
  \institution{Apple}
  \city{Cupertino}
  \state{CA}
  \country{USA}
}
\email{cforet@apple.com}

\author{Qi Shan}
\orcid{1234-5678-9012}
\affiliation{%
  \institution{Apple}
  \city{Seattle}
  \state{WA}
  \country{USA}
}
\email{qshan@apple.com}

\author{Xiaoyi Zhang}
\orcid{1234-5678-9012}
\affiliation{%
  \institution{Apple}
  \city{Seattle}
  \state{WA}
  \country{USA}
}
\email{xiaoyiz@apple.com}

\renewcommand{\shortauthors}{Hohman, et al.}

\begin{abstract}
On-device machine learning (ML) moves computation from the cloud to personal devices, protecting user privacy and enabling intelligent user experiences.
However, fitting models on devices with limited resources presents a major technical challenge: practitioners need to optimize models and balance hardware metrics such as model size, latency, and power.
To help practitioners create \textit{efficient ML models}, we designed and developed \system{}: a model visualization and optimization system.
\system{} enables practitioners to compile models to hardware, interactively visualize model statistics, and simulate optimizations to test the impact on inference metrics.
Since its internal deployment two years ago, we have evaluated \system{} using three methodologies: (1) a log analysis highlighting its growth of 800+ practitioners submitting 3,600+ models; (2) a usability survey with \NumSurveyParticipants users assessing the utility of 20 \system{} features; and (3) a qualitative interview with the \NumInterviewParticipants most active users about their experience using \system{}.

\end{abstract}



\begin{teaserfigure}
  \includegraphics[width=0.92\textwidth,alt={A screenshot of the \system{} user interface. The interface is roughly split in half, where the left side shows a rich data table of statistics about a model, and the left half shows a network graph diagram of the model.}]{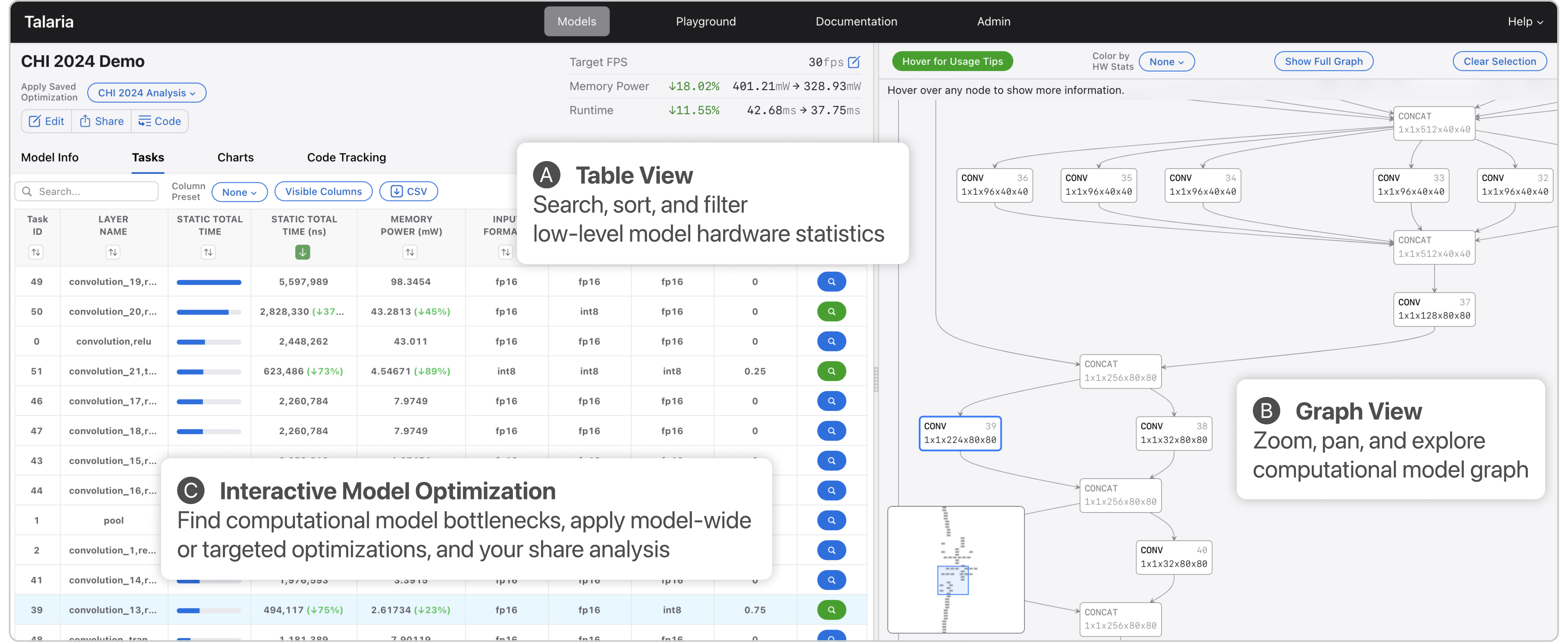}
  \centering
    \caption{
      \system{} enables ML practitioners to compile models to hardware, jointly visualize their operations in the \textbf{(A) \tableview{}} and \textbf{(B) \graphview{}}, while simulating a suite of \textbf{(C) Interactive Model Optimization{}} options to improve hardware inference efficiency.
      In this example, a user has sorted the operations by their compute time, selected one (highlighted in blue in both the table and graph), and applied an optimization that saves 18.02\% memory power and 11.55\% runtime latency.
  }
  \label{fig:teaser}
\end{teaserfigure}

\maketitle

\section{Introduction}
\label{sec:introduction}

A continuing trend within machine learning (ML) research and development is to move inference computation away from cloud servers and instead on to personal computing~\cite{apple2021ondevice, apple2022multi, apple2022deploying} and edge devices~\cite{li2018learning}.
Commonly referred to as on-device ML~\cite{google2022why}, or colloquially tinyML~\cite{warden2019tinyml}, this approach: (1) protects user privacy since data does not leave a user's device when computing inference, (2) enables new user experiences, especially for applications with strict latency requirements (\eg inference at high refresh rates), (3) supports more portable experiences since models do not require internet access, and (4) allows developers without extensive compute resources to deliver ML experiences, reducing cost and the environmental impact of large servers.
However, as the latest ML models continue to grow in size (\eg neural networks with hundreds of billions of parameters~\cite{villalobos2022machine, owid2022artificialintelligence, zhao2023survey, stanford2023ai}), creating \textit{efficient ML models} that can run inference on resource-constrained devices, such as phones, tablets, or wearables, is challenging, as deployment requires practitioners to optimize and compress their models while maintaining acceptable accuracy~\cite{vasu2022improved}.

Besides model quality metrics (\eg accuracy), how do ML practitioners effectively optimize and balance on-device inference efficiency, such as model size, power, and latency~\cite{hohman2024compression,banbury2020benchmarking}?
Efficient ML research and development is still nascent, and the state-of-the-art is rapidly changing~\cite{zhao2022survey, gu2021server, zamzam2019resource, dhar2021survey, sehgal2019guidelines}.
Best practices are largely undocumented or still forming~\cite{nvidia2023docs, warden2019tinyml}.
Much of the progress in efficient ML focuses on contributing novel compression algorithms---unfortunately much less work focuses on developing practical tools to help people successfully apply and understand the benefits of compression.
As efficient ML techniques are driven forward by advances in hardware engineering and ML research, there remains a major barrier in helping ML practitioners apply these techniques for designing real-world and intelligent ML user experiences.

The tooling for developing efficient ML models is underexplored, underdeveloped, yet rich with opportunity~\cite{hohman2024compression}.
In this timely area, better tools can have an outsized impact.
Tooling for ML is often a force multiplier, enabling practitioners of varying expertise to develop models on their own.
Interactive tools for model optimization and compression is a new direction of research, where the few existing works only scratch the surface.
Beyond communicating the effect of applying specific algorithmic compression techniques~\cite{li2020cnnpruner,dotter2018visualizing,xie2017visualization}, there are many other components of efficient ML development where interactive visualization could help practitioners create ML-powered, on-device user experiences.

To help ML practitioners build efficient models, we designed and developed \system{}: a model optimization and visualization system, informed by and built with expert ML practitioners at \location that specialize in developing efficient models on-device.
\system{} compiles models to hardware, and visualizes low-level hardware and model statistics through a split interface showing an interactive table and model graph, as shown in \Cref{fig:teaser}.
\system{} also simulates a suite of model optimizations to instantly show the impact on a model's inference efficiency (\eg latency and memory).
ML practitioners can apply these optimizations at the model level, or at the individual hardware operation level.
The system is model agnostic and supports models for arbitrary ML tasks, such as vision (\eg classification, object detection, segmentation), natural language processing, and sensing applications.

As the field of efficient ML matures, we expect model evaluation tooling to support practitioners in optimizing their models over both model behavioral metrics (\eg accuracy, precision, recall) as well as hardware specific metrics (\eg model size, latency, power consumption).
However, everything comes at a cost, and in ML, the CACE principle~\cite{sculley2014machine}, \textit{``Changing Anything Changes Everything,''} continues to hold.
Shrinking a model to reduce its size, latency, and power, while maintaining its accuracy and quality is extremely challenging in practice~\cite{hohman2024compression}.
In this work, we intentionally focus on the new and novel challenges brought by moving ML inference onto personal computing devices for enabling user experiences powered by ML.
Therefore, \system{} is scoped to help practitioners address evaluating a model's hardware metrics under the task of \textit{on-device inference} (further discussed in \Cref{subsec:inference}).

We developed \system{} over 2 years, and report on 3 evaluations.
First, we present a log analysis showing \system{}'s successful adoption within our organization.
Next, we discuss the results from a usability survey with \NumSurveyParticipants ML practitioners where they rate the utility of 20 different system features.
Lastly, we detail the results from qualitative interviews with the \NumInterviewParticipants most active users to learn about their experience using \system{} and what improvements could be made to better help them create efficient models.

Our contributions include:
\begin{itemize}

    \item \textbf{Formative research with \NumFormative ML practitioners on model optimization.}
    Through a needfinding survey and participatory design sessions with low-fidelity prototyping, we outline the challenges and tasks of optimizing a model's power consumption, memory footprint, and inference latency in order to create efficient ML models.
    
    \item \textbf{\system{}}\textbf{: an interactive visualization system for creating efficient ML models.}
    \system{} compiles models to hardware, visualizes their low-level statistics and computational graph together, while simulating multiple model optimizations for testing inference efficiency (\eg latency and memory).
    The web-based system allows users to interact with large models (\eg thousands of operations) in real time.
    \system{} also introduces a mechanism to map hardware operations back to a model's source code.
    Lastly, the system supports collaborative model optimization by letting users save optimizations and send a single URL to their colleagues to fork and continue their work.
    
    \item \textbf{Findings from three evaluations of \system{} deployed within ML research and development teams.}
    We conduct a log analysis to inspect the adoption of our system over time (800+ unique users uploaded over 3,600 models), a usability survey with \NumSurveyParticipants ML practitioners to rate and assess the utility of 20 system features, and a semi-structured qualitative interview with the \NumInterviewParticipants most active users to learn about their experience using \system{} for model optimization.

\end{itemize}

We believe efficient ML, specifically for on-device use cases, is a rich and untapped area of AI/ML for the human-computer interaction community to engage with.
There is a large gap between current tools today and what practitioners need.
We hope our work emphasizes the need and importance of tooling for optimizing models, and inspires future interdisciplinary work on interactive interfaces for creating intelligent and efficient ML user experiences.
\section{Background and Related Work}
\label{sec:related-work}

\subsection{Model Compression Techniques}
\label{subsec:background}
\begin{figure}[tb]
 \centering
 
 \includegraphics[width=\columnwidth,alt={Visual analogies for how three compression techniques work. Each start with an original apple emoji and show the output of the compression technique after it is applied. In quantization, the apple emoji becomes blurry with noticeable artifacting around the edges due to a decrease in resolution. In pruning/sparsification, the apple emoji becomes masked with a grid of white squares, where the apple is still visible but parts of it have been removed. In palettization, the apple emoji's colors are reduced to a set of six colors.}]{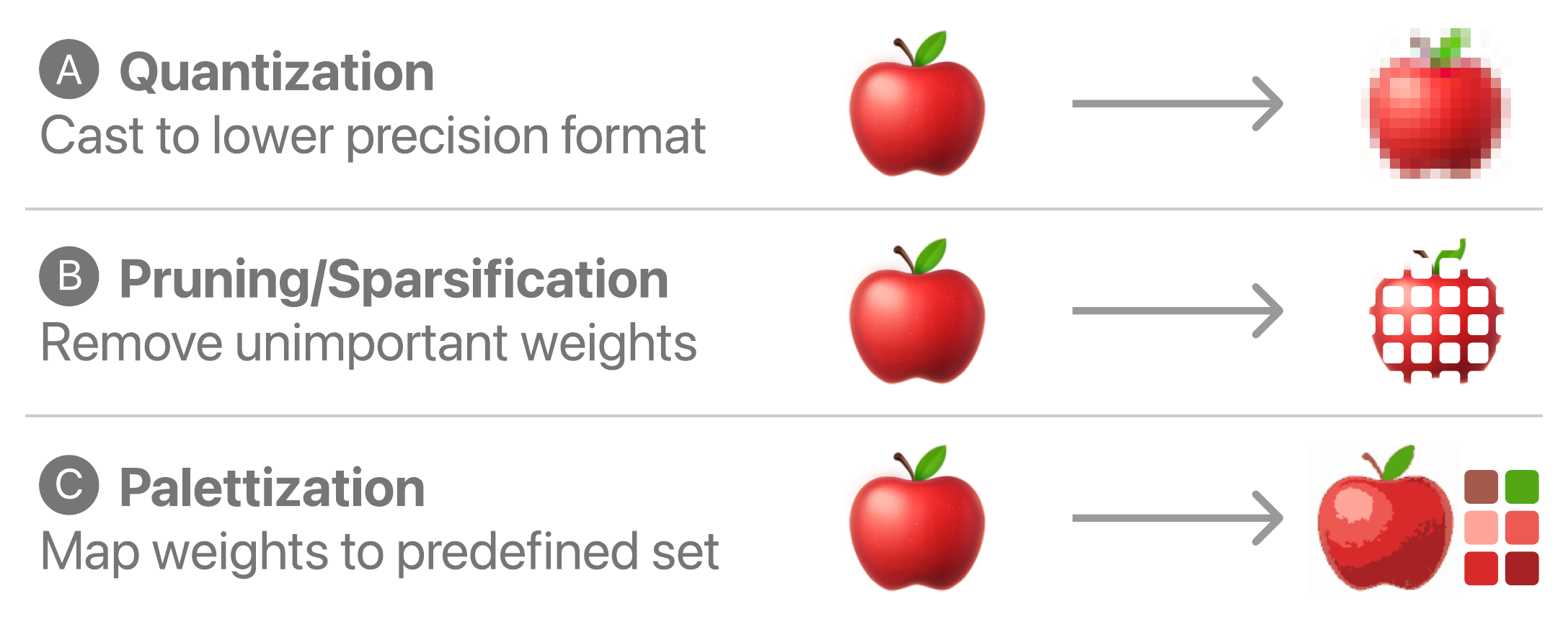}

 \caption{
 An illustration of three common model compression techniques built into \system{}.
 \textbf{(A) Quantization} converts data types from high-precision formats (\eg \texttt{fp32}) to low-precision formats (\eg \texttt{int8}).
 \textbf{(B) Pruning/Sparsification} removes unnecessary weights from neural networks.
 \textbf{(C) Palettization} maps model weights to a discrete set of precomputed (or learned) values.
 }
 \label{fig:compression-techniques}
\end{figure}

To shrink models, efficient ML practitioners use a variety of strategies, from principled architecture decisions to ad-hoc tricks-of-the-trade.
One class of techniques is model compression: optimizations to various components of a model to minimize the amount of computational resources it needs.
Categories of compression techniques (illustrated in \Cref{fig:compression-techniques}) include quantization~\cite{gholami2021survey}, palettization~\cite{cho2022differentiable, wu2018deep}, pruning (\ie sparsification~\cite{hoefler2021sparsity,gholami2021survey}), and other modeling specific techniques (\eg distillation~\cite{gou2021knowledge,gholami2021survey,polino2018model}, efficient neural architectures~\cite{vasu2023fastvit,howard2017mobilenets,sandler2018mobilenetv2,zhang2018shufflenet,tan2019efficientnet,choudhary2020comprehensive}, and dynamic architectures~\cite{zhu2021dynamic}).
Each technique is truly a family of techniques, with many nuanced variations that can be also combined together~\cite{han2015deep}.
The following surveys detail compression techniques:~\cite{menghani2023efficient, deng2020model, choudhary2020comprehensive, cheng2018model}.

In this work, the compression techniques we use are quantization (\Cref{fig:compression-techniques}A), pruning (\Cref{fig:compression-techniques}B), and palettization (\Cref{fig:compression-techniques}C).
For brief context, \textit{quantization} converts the inputs, outputs, and/or weights of a model from high-precision formats (\eg \texttt{fp32}) to lower-precision formats (\eg \texttt{fp16}, \texttt{int8}, and even \texttt{int2}) ~\cite{gholami2021survey}.
\textit{Weight pruning} removes the least-important parameters (\eg weights, bias) of a model to make it smaller.
The motivation is that modern neural networks are overparameterized, such that removing parameters will minimally impact the final prediction~\cite{hoefler2021sparsity,gholami2021survey}.
Lastly, \textit{palettization} maps the weights of a model to a discrete set of precomputed (or learned) values. 
Inspired by an artist's ``palette,'' the idea is to map many similar values to one average or approximate value, then use those new values for computing inference.
While there are many types of compression techniques, we focus on these three due to their popularity, performance, and common use.

\subsection{Existing Compression Resources}
\label{subsec:existing}
Since investing in model compression is typically only needed for applications where models will run on-device, research and best practices for ML optimization is much more limited compared to ML in general.
While surveys detail different compression techniques~\cite{deng2020model, choudhary2020comprehensive, cheng2018model}, most existing practical guidance stems from online tutorials and documentation from popular ML libraries.
Examples include TensorFlow's model optimization toolkit and blog post on quantization-aware training~\cite{tf2020quantization,tf2018introducing}; PyTorch's experimental support for quantization~\cite{pytorch2023quantization}, sparsity~\cite{pytorch2023sparsity}, and it's accompanying examples~\cite{pytorch2023examples}; Google's quantization extension to Keras called QKeras~\cite{qkeras}; Microsoft's Neural Network Intelligence package and tool~\cite{ms2021nni}; Intel's Neural Compressor library~\cite{intel2020nc}; and Apple's MLX framework~\cite{apple2023mlx} and DNIKit~\cite{welsh2023dnikit}.
For targeting specific hardware, other examples focus on speeding up inference on FPGAs~\cite{fahim2021hls4ml} and compressing Core ML models to run on Apple platforms~\cite{coremltools}.
Lastly, the appropriately named TinyML community has emerged around this topic, which published a book~\cite{warden2019tinyml} on developing models for always-on, low-power use cases.

\subsection{On-device Inference v. On-device Training}
\label{subsec:inference}
It is important to clarify a distinction between on-device \textit{inference} and on-device \textit{training}.
In our work, we focus on the more commonly studied and applied component of on-device inference: computing a prediction from a pretrained model loaded on a device with limited compute resources, such as a phone, tablet, or wearable, that have smaller memory and power capacity~\cite{hohman2024compression}.
On these specific mobile computing devices, it is rare to train a model from scratch.
In some ML contexts where personalization is needed, perhaps a model requires fine-tuning on a user's data on-device; however, this scenario is much less common than training a model offline and deploying it onto a mobile device to run inference~\cite{hohman2024compression}.
While on-device inference and training share many similar challenges, and both could benefit from interactive tools and visualization, training on-device models is not as commonplace and usually requires more resources~\cite{zhou2019edge}.
Thus, we intentionally scope our work on building tools for model optimization for ML that will run on-device inference.
For resources on the current research and challenges around on-device learning instead, see the following surveys:~\cite{dhar2021survey, lim2020federated, zhou2019edge, murshed2021machine}.

\subsection{Visualization for Model Evaluation}
\label{subsec:vis-for-eval}
Since the boom of ML innovation over a decade ago, there have been many visual analytics systems designed for most stages of the ML development cycle.
This hybrid research direction of combining visualization and ML has made significant contributions to model evaluation~\cite{hohman2018visual}.
For different modeling tasks, tools for visualizing metrics (\eg accuracy, precision, recall) on subsets of data~\cite{cabrera2023zeno, cabrera2019fairvis, ahn2019fairsight, wexler2019if} and tools for exploring large ML datasets~\cite{bertucci2022dendromap, kyd} help practitioners compare and evaluate how well ML models generalize to unseen data.
Example ML tasks incorporating visualization include data classification~\cite{amershi2015modeltracker, ren2016squares, goertler2022neo}, image classification~\cite{choo2010ivisclassifier}, object detection~\cite{gou2020vatld}, transfer learning~\cite{ma2020visual}, and natural language processing (NLP)~\cite{strobelt2022interactive, strobelt2017lstmvis, hoover2019exbert, brath2023role}.

Research into how ML practitioners build and evaluate models in code has shown that ML code is highly experimental and iterative compared to conventional programming~\cite{patel2008investigating, amershi2019software}.
This observation has generated new ways of incorporating visualization into ML development processes, \eg enhancing computational notebooks~\cite{bauerle2022symphony,kery2020mage}.
However, for all the emphasis on evaluating model behavior, there are much fewer visualization tools that evaluate a model's efficiency (\eg latency and power consumption).
The few tools that exist show model metrics, but do not inform ML practitioners of the potential efficiency improvements from the latest model optimization and compression techniques.

\subsection{Visualization for Model Optimization}
\label{subsec:vis-for-opt}
Compared to general model evaluation, there are few existing visualization tools for efficient ML optimization.
Most work studies and surveys algorithmic techniques to compress models, such as sparsification~\cite{hoefler2021sparsity}.
Tooling is much less developed~\cite{hohman2024compression}.
One of the few related visualization works to ours is CNNPruner~\cite{li2020cnnpruner}, which focuses on one specific compression technique, pruning, for convolutional neural network architectures.
Other work shows only static visualizations of results and features during model optimization; for example, \citet{dotter2018visualizing} analyzed model metrics such as inference time and model size along with visualizing data clusters for a classification task, and \citet{xie2017visualization} visualized features learned by a network as guidance to better prune redundant kernels.
Model graph visualizers, such as the TensorFlow Dataflow Graph visualizer~\cite{wongsuphasawat2018visualizing} and open-source tools like Netron~\cite{lutz2017netron}, allow practitioners to inspect their models, but are not designed for the task of optimization.
Most existing tools are not grounded in real-world workflows and needs of ML practitioners, nor do they factor in details about a model's efficiency and hardware metrics.

\section{Formative Research: Motivation and Challenges}
\label{sec:design-challenges}

From literature it is clear that tooling for creating efficient ML models is underdeveloped.
This is in part due to the specialized nature of on-device ML: building optimized models brings all the challenges of conventional ML development, but additionally requires niche expertise in hardware knowledge and access~\cite{hohman2024compression}.

Motivated by these challenges, we sought to explore opportunities where visualization could help.
To build the right tools for model optimization, we conducted formative research to better understand the challenges and needs for creating efficient models.
We first conducted a small needfinding survey with ML practitioners at \location (\Cref{subsec:formative-survey}).
Then through participatory design sessions, we developed low-fidelity prototypes on practitioner data to engage them with what interactive visualization could offer (\Cref{subsec:formative-prototype}).

\begin{table}
\centering
\caption{A summary of the completed responses to the needfinding survey, including their role, primary type of ML application, and years of experience in ML.}
\label{tab:survey-participants}
\begin{tabular}{lllc}
\textbf{ID} & \textbf{Role} & \textbf{ML Application} & \textbf{Exp.}\\
\midrule
P1 & ML Manager & Deployment \& Optimization & 10 \\
P2 & ML Engineer & Training \& Optimization & 9 \\
P3 & ML Engineer & Training \& Optimization & 8 \\
P4 & Research Scientist & Research \& Optimization & 9 \\
P5 & ML Engineer & Training \& Optimization & 5 \\
P6 & ML Engineer & Training \& Optimization & 4 \\
P7 & ML Engineer & Deployment \& Optimization & 4 \\
P8 & Research Scientist & Research \& Optimization & 5 \\
P9 & ML Manager & Training \& Optimization & 7 \\
P10 & ML Engineer & Training \& Optimization & 3 \\
P11 & Research Scientist & Research \& Optimization & 6 \\
P12 & ML Engineer & Deployment \& Optimization & 5 \\
\end{tabular}
\end{table}

\subsection{Needfinding Survey for Efficient ML}
\label{subsec:formative-survey}

To begin, we sent out an open-ended needfinding survey to efficient ML experts within our organization to ask what features interactive tools for model optimization should support.
The survey format consisted primarily of open-ended text responses and was largely unstructured to gather diverse perspectives on optimizing models.
We received \NumFormative responses, summarized in \Cref{tab:survey-participants}.
The participant count of our survey is lower than others within our organization because we made participation criteria strict: participants were required to be experts in efficient ML, hardware optimization, and at least one area of ML (\eg research, model training, or deployment), to ensure the data was as relevant and informed as possible.
With \NumFormative participants, they had 75 years of experience between them.
We note that this survey was conducted solely within one organization, therefore practitioners may hold organization-specific beliefs and practices~\cite{schein1990organizational}.
However, between existing field studies on efficient ML in practice~\cite{hohman2024compression}, the number of years of experience, and the specialized expertise shared by these participants, we are confident that our findings accurately describe current challenges within their work, and efficient ML more broadly.

With regards to what features new tooling could support, many requests were domain specific to ML model and hardware analysis, such as attributing power and memory consumption to individual ML operations executed on-device.
All \NumFormative responses (P1--P12) indicated a specific metric that they regularly inspect (\eg model size, inference speed, memory usage, memory power).
Analyzing these statistics is one of the primary routine analyses efficient ML practitioners perform.
Therefore, the ability to extract these statistics from an arbitrary model and quickly load them into tools for analysis will shorten the time it takes for practitioners to visualize and optimize their models.
Responses made it clear that for any tool to be successful in this work, it must support this task.

However, responses indicated that only analyzing the model and hardware statistics is not enough; ML practitioners also need to know the locations of these metrics inside models (\ie geometrically within the compiled computational graph).
Practitioners do not only want to know in aggregate how much computational budget (\ie a threshold for model size, latency, power or an amount of any specific resource a model is allowed to consume) their models use, but they additionally want to know specific operations within the model these aggregates are heavily weighted from.
Nine responses (P1--P6, P9--P11) expressed their desire for tools to help them sort, filter, and locate the biggest ``offenders'' (the most computationally expensive operations).
Also referred to as computational bottlenecks, these are high-value hardware operations that help practitioners minimally edit models.
Since it becomes harder to have an accurate model the more optimization is applied, leaving as much of the original model intact is a desirable approach.
Computational bottlenecks in this case are prime candidates for potential optimization savings that practitioners want to know about.

Another group of eight responses (P2--P7, P9, P10) expressed enthusiasm for quickly testing optimization options to see the impact on hardware metrics.
Quick optimization experimentation is important, as different optimizations will have different effects on the model's metrics, and it can be hard to know what the effect of optimizing a single layer will be to the entire model.
Lastly, a common theme was the inherent collaborative nature of this type of work: it requires not only ML engineers, but also hardware specialists, compiler engineers, and people with hybrid expertise who can float between these roles.
These practitioners have a niche, but high-demand and hybrid skillset that cannot scale with the amount of projects they work on.
Tools that help them analyze models more quickly, share the results (\eg overall latency improvements, layer-level memory analyses, or the impact of optimization before and after its applied), and perhaps educate other ML engineers about optimization techniques can help distribute their expertise.

\subsection{Participatory Design and Low-Fidelity Visualization Prototyping}
\label{subsec:formative-prototype}

Given the perspectives we found from the needfinding survey, we next wanted to gather more insight into creating efficient models by letting the survey participants interact with basic prototypes.
After obtaining data from one in-development model, we built low-fidelity prototypes and visualizations to provide the ML practitioners with tangible artifacts to inspect and critique.
To gather the most precise and informative qualitative feedback, it was important to prototype with real data and models.

Over the course of a month, we met weekly with the \NumFormative participants, updating our prototypes based on both their requests and our expectation on useful features.
These prototypes were often specific yet disjoint solutions to problems raised in the needfinding survey.
For example, one prototype was a rich data table that showed all the different metrics that could be gathered from a model compiled to run on hardware.
The practitioners (P1--P12) said this was a must-have, and appreciated quickly sorting and filtering operations to find model bottlenecks and more generally see the overall distribution of compute used within the model.
This first table prototype was a direct result of the needfinding survey task where practitioners all mentioned specific metrics they wanted to gather and analyze together, as oftentimes they are making trade-offs between multiple metrics (\eg does making the model faster in one location increase its memory usage?).
Later on we added results from precomputed optimizations on the model as well, which practitioners (P2-P7, P9--P11) said was helpful in having optimized model data alongside the original model.

Another prototype was a simple dashboard that implemented basic interactive visualization techniques (\eg brushing and linking, details on demand).
Practitioners (P1--P3, P8, P11) appreciated this alternative, visual view of the data from the table, but said that they constantly are inspecting specific operation values, so the table should almost always be on screen.
This dashboard prototype was then positioned as complementary.

One other prototype was a simple node-link diagram of a model's hardware operations.
Practitioners (P1--P9) greatly appreciated seeing the structure of a model.
We then added controls to encode nodes of the graph by different metrics to highlight where in the model certain metrics were heavily weighted.
This was illuminating to the practitioners, as they had not produced a visualization like this before, but have always wanted a view to find bottleneck operations geometrically in the model, not only from statistics.

By the end of the month, we had a small collection of prototypes, ranging from data tables, dashboards, computational graphs, and others, that was sufficient for demonstrating power of interactive visualization in efficient ML development.
When reviewing all the prototypes with the practitioners, they again stressed inspecting their models analytically and geometrically, and that each view gives a different perspective to their work.
It was agreed upon that the foundation of a future tool should support both paradigms.
These prototypes helped prioritize system capabilities during our design and development of \system{}.

\subsection{Design Challenges for Model Optimization}

From combining the data gathered from our needfinding survey (\Cref{subsec:formative-survey}) and feedback from the low-fidelity visualization prototypes (\Cref{subsec:formative-prototype}), the most common and pressing challenges for optimizing ML models coalesced, which we list as (\textbf{C1--C5}) below.

\begin{itemize}

    \item[\textbf{C1.}]
    \textbf{Inspecting model statistics analytically and geometrically.}
    Efficient ML analysis requires looking at both large amounts of tabular model statistics and large network diagrams simultaneously.
    It is time consuming and cumbersome, yet critical, to toggle back and forth between these two views.
    
    \item[\textbf{C2.}]
    \textbf{Finding model bottlenecks.}
    Not every piece of a model needs to be, or should be, optimized.
    It is hard to find computational model bottlenecks and place them in context with the global architecture.
    
    \item[\textbf{C3.}]
    \textbf{Interactively testing multiple model optimizations.}
    Tools for model compression are in their infancy, and lack interactive interfaces to support general optimization analysis.
    It is unclear to know how much and where to apply model optimizations to hit target metrics and computational budgets.
    
    \item[\textbf{C4.}]
    \textbf{Collaboratively optimizing a model.}
    Efficient ML work requires multiple practitioners and experts to iteratively make decisions during model development.
    It is difficult to keep track of shared analyses from multiple contributors.
    
    \item[\textbf{C5.}]
    \textbf{Accurately applying model optimizations.}
    Translating findings from optimization analyses into practice (\eg applying compression to a layer in a model's training code) can be time consuming and error prone.
    
\end{itemize}

\section{Visualization System Requirements and Task Analysis}
\label{sec:tasks}

From our formative research, there is clear opportunity to help practitioners create efficient ML models.
Practitioners reported that existing tools were insufficient, and expressed enthusiasm that visualization could help them develop smaller, more efficient models for on-device user experiences.
Given the relatively novel domain and sparsity of work that addresses this budding area of ML, we sought to design new interactive visualizations for optimizing ML models.
To inform our design, we distilled five main tasks performed by practitioners that our system should support. 
The tasks (\textbf{T1–T5}) below are mapped to the challenges (\textbf{C1–C5}) raised in \Cref{sec:design-challenges}:

\begin{itemize}

    \item[\textbf{T1.}] Quickly analyze low-level model and hardware statistics to understand a model's inference (in)efficiency (\textbf{C1, C2}).
    
    \item[\textbf{T2.}] Interactively visualize model architecture to see its topology and to find computational performance bottlenecks in the computational graph (\textbf{C1, C2}).
    
    \item[\textbf{T3.}] Explore varying model optimizations and quickly examine their effect on inference efficiency, including both model-wide and targeted optimizations (\textbf{C3}).
    
    \item[\textbf{T4.}] Allow teams to collaboratively optimize models (\textbf{C4}).
    
    \item[\textbf{T5.}] Make optimizations actionable by attributing low-level hardware operations to their source code locations to help practitioners know where to implement optimizations (\textbf{C5}). 
\end{itemize}

\section{\system{} Interface and System}
\label{sec:system}

With the tasks identified from our formative research, we
present \system{}, an interactive visualization for ML model optimization.
\system{} enables ML practitioners to understand how their models perform on-device and optimize them for improved inference efficiency.
The system visualizes hardware statistics through a split interface showing an interactive table and model graph.
\system{} is a substantive engineering effort, containing many features that address challenges practitioners face when building efficient ML.
The system is model agnostic and supports arbitrary ML tasks, such as vision, NLP, and sensing. 
Throughout this section, we link relevant views and features to the tasks (\textbf{T1–T5}) identified from our task analysis (\Cref{sec:tasks}).

\subsubsection{System Header}
The \system{} header contains top-level information about a model, including key statistics that practitioners need to know and optimize, such as the targeted inference frame rate (fps), memory power (mW), and latency (ms).
The header also contains the main navigation tabs for \system{}, to switch between the specific visualizations and views described below.
When switching views, the system header remains fixed in the interface.

\subsection{The \tableview{}}
\label{subsec:system-table}

The first main view of the interface is the \tableview{} (\Cref{fig:teaser}A), a rich, interactive data table that displays the low-level hardware statistics of how a model will run (\textbf{T1}).
Each row of the table corresponds to one low-level hardware task, and each column encodes different metrics. 
One important metric is the clock time it took for a task to run (\texttt{TOTAL TIME} column), which is dual encoded in this table as both a number and an inline sparkbar~\cite{tufte1986visual}.

There are dozens of metrics to visualize, but the system displays only a few by default; the default options were chosen based on practitioners' feedback from the formative research in \Cref{sec:design-challenges}.
Users can add, remove, or browse all the available metrics by clicking the ``Visible Columns'' button.
Users who are not familiar with each metric can hover over the metric name in the column header to display a tooltip that describes the metric in plain language.

The \tableview{} also supports common tasks for interacting with rich data tables that practitioners requested from our participatory design sessions.
Users can sort the table by a metric when they click the arrow icon in a column header, filter the table (\eg show tasks that took longer than 1ms), and search by the task name or ID. 
These features allow users to quickly explore and analyze the statistics of their models. 

Lastly, the \tableview{} is interactively linked to the \graphview{}.
For example, selecting a task in the table will zoom in and highlight the correspond node in the graph.
This is a simple but critical interaction, as it allows practitioners to link task statistics to their location in the model's graph for further analysis.
Multiple selections are also supported, \eg when the table is filtered to a subset of tasks, the \graphview{} highlights the selected task and auto-resizes the graph to show these tasks.
This shared state is a pattern within \system{}: interactions in one view are linked with the others in the system.
We decided to implement multi-coordinated views and cross-filtering from our needfinding survey since practitioners lamented that they frequently toggle back and forth between statistics and graph visualizations.

\subsection{The \graphview{}}
\label{subsec:system-graph}

\begin{figure}[tb]
 \centering

 \includegraphics[width=\columnwidth,alt={Five neural network models that are represented as node-link diagrams. The diagrams are arranged from left to right, where each becomes larger and more complex, \eg, more nodes and more edges.}]{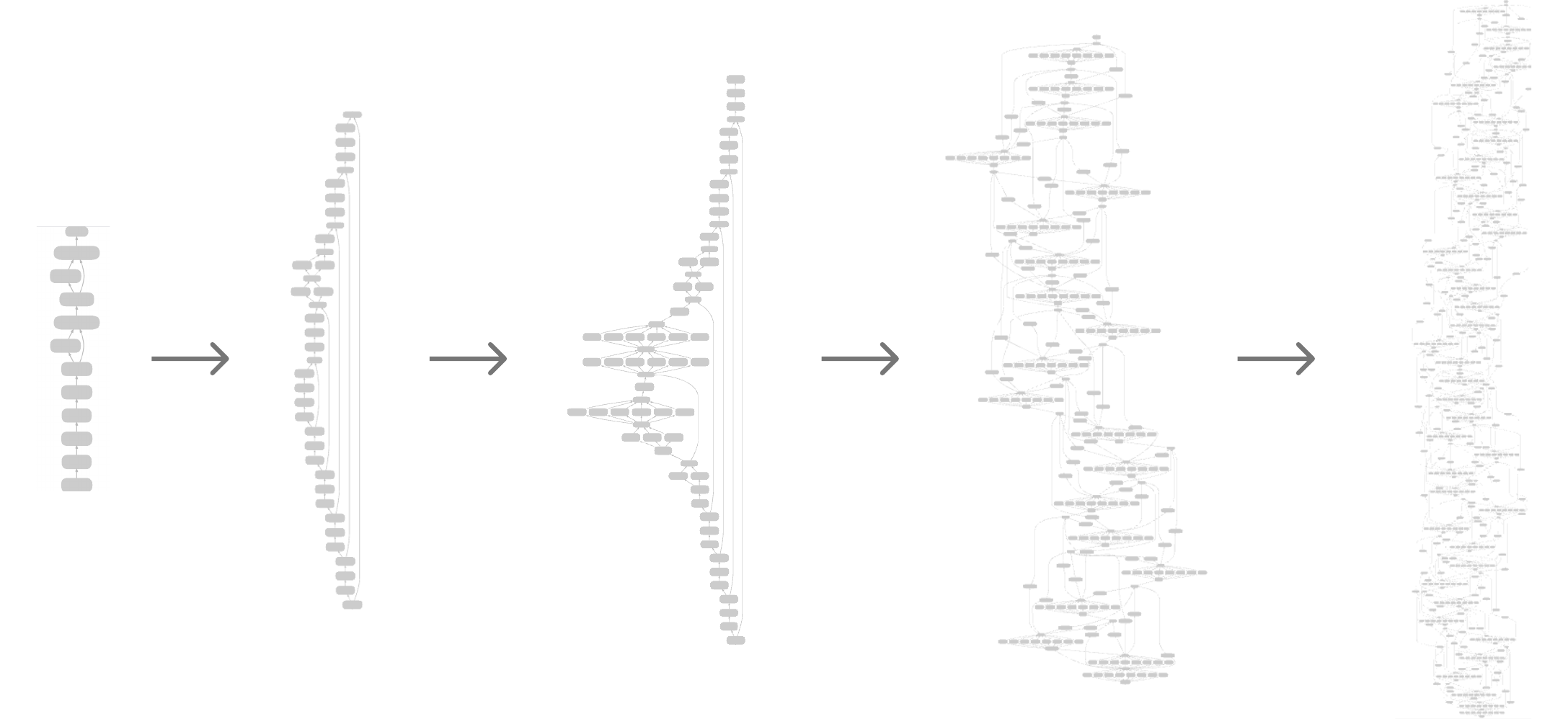}

 \caption{
 Five different models visualized in \system{} with increasingly complex architectures.
 }
 \label{fig:mini-graphs}
\end{figure}

\begin{figure}[!b]
 \centering

 \includegraphics[width=\columnwidth,alt={A neural network model represented as a node-link diagram, where the nodes are colored shades of blue indicated which node has a high value of some metric of interest. There are three copies of the same network side by side, each with different patterns of shaded blue nodes to compared different metrics within a model.}]{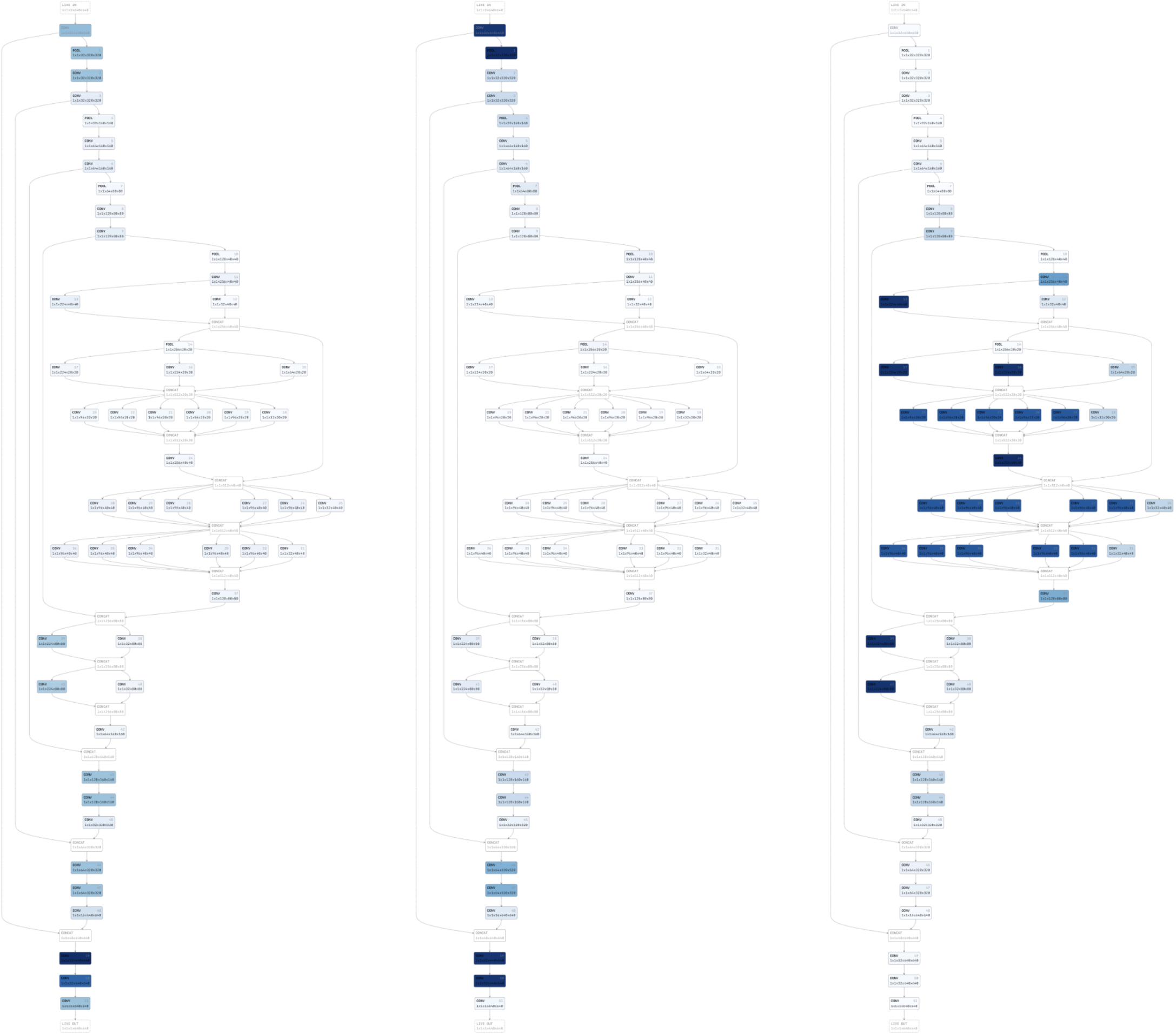}

 \caption{
 Three examples of the \graphview{} encoding different hardware metrics on the same model to quickly identify potential model bottlenecks.
 Dark blue nodes indicate higher values for a metric, \eg latency, memory, or power usage.
 }
 \label{fig:graph}
\end{figure}

The second main view of the interface is the \graphview{} (\Cref{fig:teaser}B), an interactive canvas that displays the compiled model architecture graph (\textbf{T2}).
Each node in the graph corresponds to a low-level hardware task (\eg a convolution or concatenation operation).
It is important to note that this graph represents the operations of a model compiled onto hardware (similar to visualizing a dataflow graph~\cite{wongsuphasawat2018visualizing}), not just the conventional model architecture from model definition code.
The computational graph shown in \system{} is richer and often more complicated (example models growing in complexity shown in \Cref{fig:mini-graphs}).

Users can freely zoom and pan on the graph to inspect how their models get compiled to hardware.
For details on demand, hovering over any node displays a tooltip with important metrics that may interest practitioners during exploration.
When a user wants to get more information about a particular task, selecting a node also highlights the corresponding task in the \tableview{}, which contains all the other available metrics as discussed above.
Besides selecting a single node, users can also select multiple nodes with a lasso selection; this selection also filters the \tableview{} to the corresponding tasks in the selection.

Since models can be large, both in depth (\eg number of layers) and width (\eg parallel layers or branches), the \graphview{} shows a minimap (a small graph overview) to allow users to quickly identify areas of interest (\Cref{fig:teaser}B).
Minimap examples for five models with increasingly complex architectures are shown in \Cref{fig:mini-graphs}.
The minimap also helps users keep the global model geometry in mind when they are zoomed into a particular region.
Users can drag the minimap selection window to reposition the main \graphview{} (\eg quickly jump to a farther away location in the model).
The minimap can also be hidden to maximize screen space.

Another technique to wrangle large models is to group relevant tasks and construct a hierarchy when appropriate.
When practitioners export models, they can define groupings in their code (\eg group all tasks in a Transformer unit, or group tasks in a specific sub-network).
With a hierarchical graph where supernodes can be interactively expanded or collapsed (taking inspiration from~\cite{wongsuphasawat2018visualizing}), practitioners can reduce the number of nodes in their view to focus on higher-level model structure.

The last important feature of the \graphview{} is coloring the graph by a model metric. 
This is critical for quickly finding computational bottlenecks within a network.
Users can pick a metric in either of two locations: (1) the dropdown menu in the \graphview{}, or (2) the ``plot'' icon in a column header in the \tableview{}.
Either selection updates the color of the nodes, where darker blue indicates more computationally expensive tasks, as seen in \Cref{fig:graph}.
This design lets dark nodes (\ie bottleneck tasks) stand out when zooming out for an overview.

\begin{figure}[tb]
 \centering

 \includegraphics[width=\columnwidth,alt={A toy neural network model is represented as a node-link diagram is gray. It has two arrows pointing if different directions. The first direction demonstrates model-wide optimization, where the entire network is now colored blue to show that this technique impacts every node and edge of the models. The second direction demonstrates targeted optimization, where only a couple nodes and edges are colored clue to show that this technique only impacts a sub-network within the model.}]{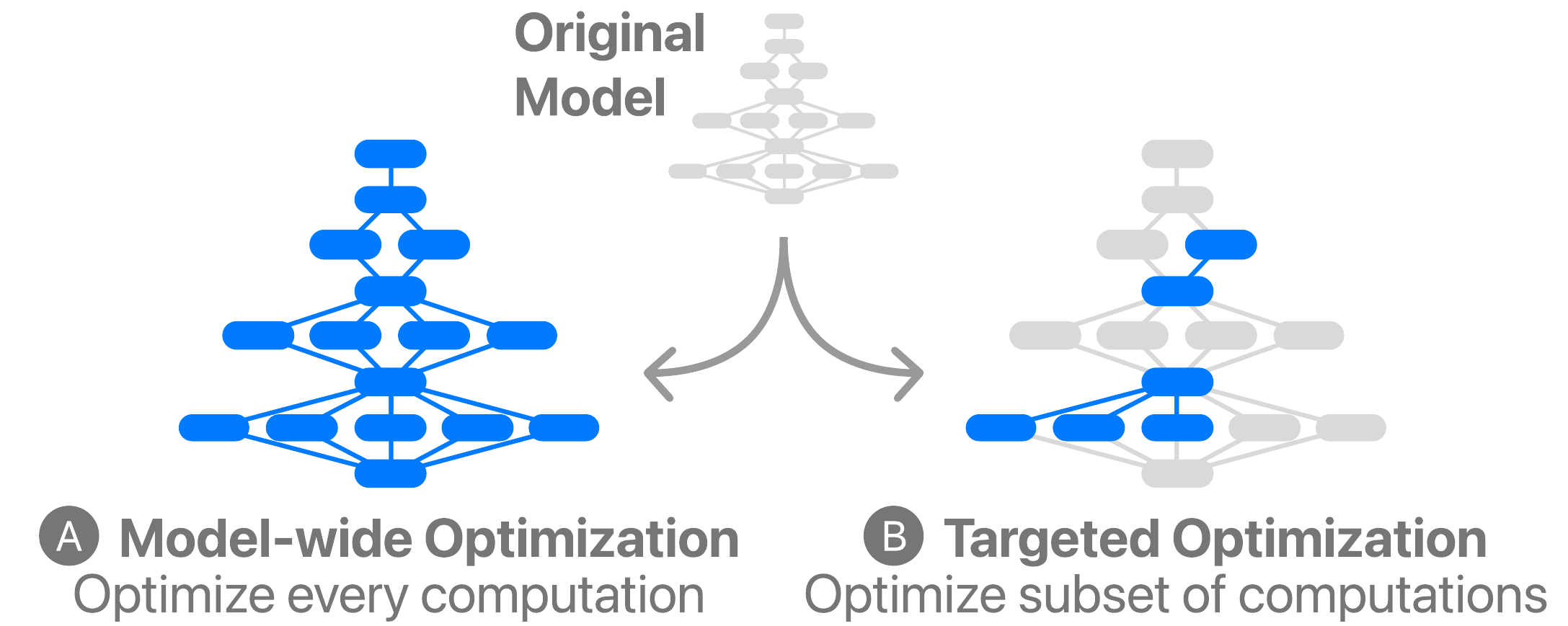}

 \caption{
 An illustration of two types of model optimization.
 \textbf{(A) Model-wide optimization} applies a compression technique to the entire model, regardless of outcome.
 \textbf{(B) Targeted optimization} only compresses certain model operations.
 \system{} supports both, and allows practitioners to interactively optimize individual model operations.
 }
 \label{fig:optimization-graphic}
\end{figure}

\subsection{Interactive Model Optimization}
\label{subsec:system-optimization}

In addition to visualizing model statistics and the compiled graph, \system{} contains powerful features to help ML practitioners make informed decisions on model optimization (\textbf{T3}).
To optimize a model, practitioners typically have to implement and apply optimizations, such as specific compression techniques, to empirically test which techniques give the best results.
This can be time consuming and feel like ``searching in the dark.''
Instead, \system{} enables users to select and compare model optimizations in real time.

How is this possible?
At compile time, \system{} precomputes many possible optimizations for every task and saves this data to the \system{} backend server.
Although these are estimations of hardware metric savings (\eg latency and power), in most of our tests, models are sufficiently accurate (within 1--3\% variance, compared to actual hardware benchmarking).
When a user selects an optimization, the interfaces updates in two places.
First, the table in the system header shows the result on the model's overall metrics (as seen in \Cref{fig:teaser}, where this optimization results in saving 18.02\% memory power and 11.55\% latency).
Second, the \tableview{} shows the new, optimized statistics for each task colored green or red depending on if they improved or regressed (\Cref{fig:teaser}).

\system{} supports two types of optimization: (1) model-wide predefined optimizations (\Cref{fig:optimization-graphic}A), and (2) task-specific targeted optimizations (\Cref{fig:optimization-graphic}B).

\subsubsection{Model-wide Predefined Optimizations}

\begin{figure}[!b]
 \centering
 
 \includegraphics[width=\columnwidth,alt={Two screenshots of the two different optimization options in \system{}. The first is for model-wide optimizations, where a short table lists out the options a user can take in natural language. The second is for targeted optimizations, where a table of different optimizations show the impact on various metrics for a single layer in the model.}]{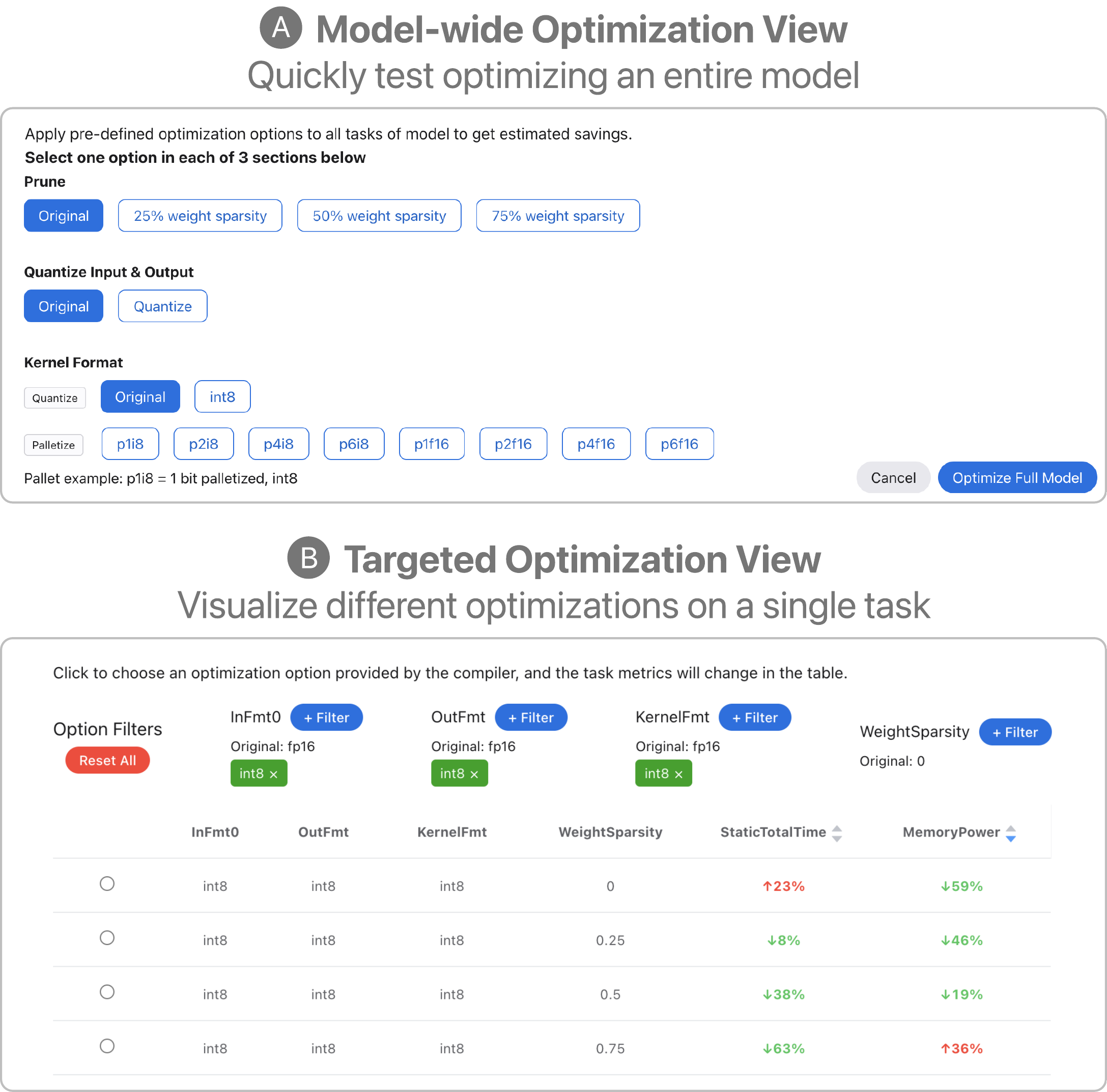}

 \caption{
 \system{}'s \textbf{(A) model-wide optimization} for quick experimentation and \textbf{(B) targeted optimization} for compressing a single hardware operation. 
 Targeted optimization displays a table where rows are different compression techniques, with metric changes colored green or red.
 In this example, a user has filtered the table to only consider optimizations where the input and output formats are quantized to \texttt{int8}.
 }
 \label{fig:optimization}
\end{figure}

Model-wide optimizations are a commonly used yet blunt approach, where the same optimization technique applies to ever single task in a model.
For example, one could either quantize or sparsify an entire network to reduce model size.
\system{} provides predefined model-wide optimizations that are most commonly considered (\Cref{fig:optimization}A).
Since \system{} allows a user to examine optimization impact in real time, this is a great first attempt when someone wants to quickly estimate latency or power savings with common model-wide optimization.

\begin{figure*}[tb]
 \centering
 \includegraphics[width=\textwidth,alt={Three complementary charts from \system{}. The first shows multiple bar charts of different metrics as histograms. The second shows a scatterplot where a rough correlation is made between the x and y axes. The third shows a waterfall chart where bars indicate how long an operation takes to run; most operations are short, whereas only a few of the model operations take up a majority of the computation time.}]{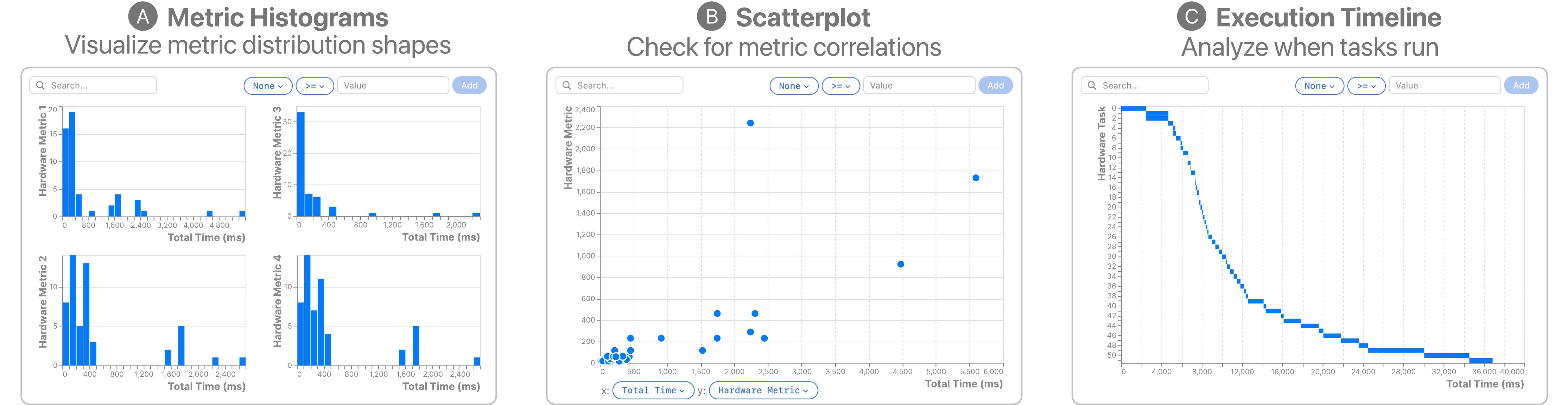}
 \caption{
 Complementary visualizations to help ML practitioners analyze their models.
 \textbf{(A) The Univariate Metric Histograms} give users a quick glance of the distribution shape of various model metrics.
 \textbf{(B) The Scatterplot} helps identify correlations between model metrics.
 \textbf{(C) The Execution Timeline} shows when the different operations of a model execute.
 }
 \label{fig:complementary-visualizations}
\end{figure*}

\subsubsection{Task-specific Targeted Optimizations}

More advanced and novel to \system{} are targeted optimizations that apply to specific tasks, for example a bottleneck task that is computationally expensive.
Whereas model-wide optimizations can be seen as coarse techniques, targeted optimizations give users fine-grain control.
Targeted optimizations avoid excessive compression of a model, which better preserves behavioral metrics like accuracy.

To optimize a task, users can click the ``Optimize'' button in the \tableview{} to see a modal that presents an exhaustive list of combinations of optimizing a task's ``Input Format'', ``Output Format'', ``Kernel Format'', and ``Weight Sparsity.''
Each optimization also shows the impact on this task's latency and memory power.
Users can filter these options to a subset of optimizations that they prefer, \eg only considering options with \texttt{int8} kernel quantization.
To help practitioners make a decision, each option's relative change among all options are colored for easier comparison.
For example, in \Cref{fig:optimization}B, green text indicates positive outcomes (\eg latency drops) and red text indicates the opposite.
While optimizing a task often leads to better inference efficiency, some optimizations make trade-offs (\eg reducing memory but increasing latency).

With the \tableview{}, the \graphview{}, and real-time optimization features, novel analysis workflows start to emerge.
ML practitioners can observe metric distribution patterns in the \tableview{}, quickly locate the model bottlenecks from the \graphview{}, then selectively optimize those tasks to squeeze out the best possible inference efficiency.
This follows a guiding design principle where practitioners want to minimal edit and optimize their models.
\system{} allows them to prioritize optimizations and get the best ``bang for buck.''

\subsection{Collaborative Optimization and Saving Compression Analyses}
\label{subsec:system-collaboration}

In practice, building ML models is a collaborative effort with multiple contributors.
\system{} was designed with this workflow in mind, and contains lightweight but important features to support collaborative model optimization for ML teams (\textbf{T4}).

A user can save an optimization in \system{} by clicking the save button and providing a name for the analysis.
An example can be seen in \Cref{fig:teaser} in the system header where a user has saved an optimization named ``CHI 2024 Analysis.''
This feature is also useful for (1) saving an analysis as a specific checkpoint, (2) tracking the path to a particular savings goal, or (3) saving an optimization and then restarting to work on an alternative.

Moreover, when a model is uploaded to \system{}, a unique URL is generated.
Once the uploader grants permission, this URL can be shared to individual users or user groups, and the model will appear in collaborators' model list page.
This is designed for a common workflow, where an ML engineer optimizes their model, saves the analysis, and sends the URL to their team for review.
Model owners can also enable link sharing, so that any other user could load a previously saved optimization, edit it, and save it as new analysis. 

\subsection{Source Code Tracking}
\label{subsec:system-code}

Once an ideal optimization is chosen, practitioners need to apply it back to their code.
\system{} supports a key feature called source code tracking which maps each hardware task back to the model definition in code (\textbf{T5}). 
To enable source code tracking, practitioners export models using \system{}'s companion framework, which constructs a graph of hardware tasks.
During graph construction, it parses the call stack of each API call to get code locations.
The exported model package includes a JSON file mapping source code to hardware tasks.
The end result is that users can trace a single task from hardware in the stack to the exact line of code of their model definition which spawned the task.
Users can interact with this feature in two views: Code Locations and the Code Browser.

\subsubsection{Code Locations View}

Selecting a task from the \tableview{} or the \graphview{} populates the Code Locations view, which shows the code snippet that spawned the task.
This allows a practitioner to quickly find which code to edit to apply the optimizations. 

\subsubsection{Code Browser View}

Each code snippet also contains the name of the file that the snippet belongs to.
Clicking on the filename changes the view to the Code Browser (a read-only, web-based code editor), which highlights the line of code from the snippet to give the practitioner better code context.
The code browser has common features of a code viewer, including a filetree browser, syntax highlighting, and a code minimap.

\subsection{Complementary Visualizations}
\label{subsec:system-complementary}

\system{} also contains three complementary visualizations to help practitioners explore model statistics.
The visualizations show model operations, \ie rows in the \tableview{} and nodes in the \graphview{}.
These views are interactive and share state within the tool, \eg selecting or filtering tasks in one view updates all other views.
Users toggle between these views from tabs in the system header.

\begin{figure}[!b]
 \centering

 \includegraphics[width=\columnwidth,alt={A diagram with nested boxes showing the relationship between the frontend, backend, database, and file storage components of the system.}]{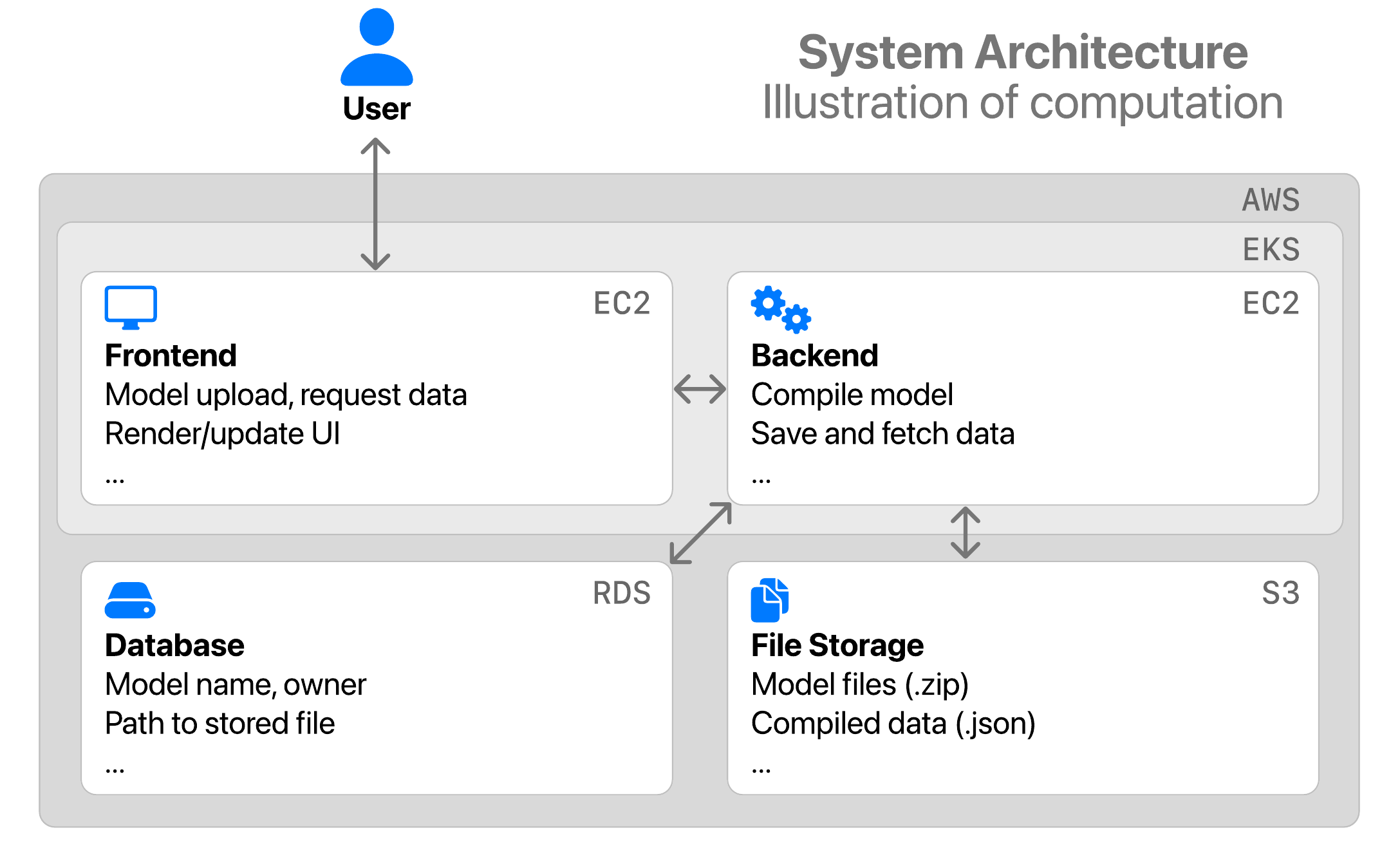}

 \caption{
 The \system{} system architecture.
 A user interacts with the web frontend to visualize the model.
 The frontend communicates with a backend server that compiles the model, and also connects to database and file storage services for saving and retrieving model information.
 }
 \label{fig:system-arch}
\end{figure}

\subsubsection{Metric Histograms}

The first complementary view is a grid of univariate histograms (\Cref{fig:complementary-visualizations}A) to give users a quick glance at the distribution shape for every metric of their model.
Lightweight interactions are available, such as a range selection to filter out parts of a distribution that are not needed; \system{} then updates the selection state of the system and remaps the axes to fit the data subset.
Filtering multiple histograms helps users find a subset of tasks that they are interested in.

\subsubsection{Scatterplot}

The second complementary view is a scatterplot (\Cref{fig:complementary-visualizations}B) that helps users find correlations between metrics.
Each axis contains a dropdown to specify a metric.
Hovering over a point displays a tooltip with task details.
Clicking or selecting points also selects those tasks in the other views of \system{}.

\subsubsection{Execution Timeline}

The third complementary view is a timeline visualization (\Cref{fig:complementary-visualizations}C) that helps users see the execution of their model's tasks chronologically.
Tasks are arranged on the y-axis, and time on the x-axis, where bars indicate how long a task took.
This encoding makes it easy to compare computationally expensive tasks (larger bars) to smaller tasks.
Moreover, this view is useful in both quickly finding top offenders, \ie computationally expensive tasks, and chronologically locating each task when it runs during inference time.
Similar to other views, clicking any task updates the \system{} selection in the other views.

\subsection{System Implementation}
\label{subsec:system-implementation}
\system{} is a web-based system built on a common web stack.
The guiding design philosophy of the system is to keep as much as the workload as possible in the browser and use a backend primarily for data and model compilation.

For the frontend, we used open-source libraries including Vue.js\footnote{https://vuejs.org/} for the primary UI framework, D3.js\footnote{https://d3js.org/} for data transformations and visualization rendering, and the Monaco Editor\footnote{https://microsoft.github.io/monaco-editor/} for displaying code.
For the backend, we used Flask\footnote{https://flask.palletsprojects.com/} as a lightweight WSGI app framework that communicates with our database and storage and serves data to the frontend.
Most of the interactivity logic is located in the frontend (\eg rendering and visualization interactivity), while the backend is mainly used to provide precomputed JSON data (\eg computing possible optimizations as mentioned in \ref{subsec:system-optimization}).
Our service is hosted on Amazon Web Services Enterprise (\eg EC2, EKS, RDS, S3)\footnote{https://aws.amazon.com/}.
For more details on how each component relates to one another, see our system architecture diagram in \Cref{fig:system-arch}.
\section{Illustrative Usage Scenario}
\label{sec:use-case}

To show how \system{}'s features described in \Cref{sec:system} work together to help ML practitioners visualize and optimize their models, we present an illustrative usage scenario.

\begin{figure*}[t]
 \centering

 \includegraphics[width=\textwidth,alt={Screenshots of the \system{} interface showing how a user \user{} meets her runtime budget. It shows a the \tableview{} of failed model-wide optimization, then the \tableview{} of a successful targeted optimization, along with the \graphview{} and code snippet of the hardware operations to optimize.}]{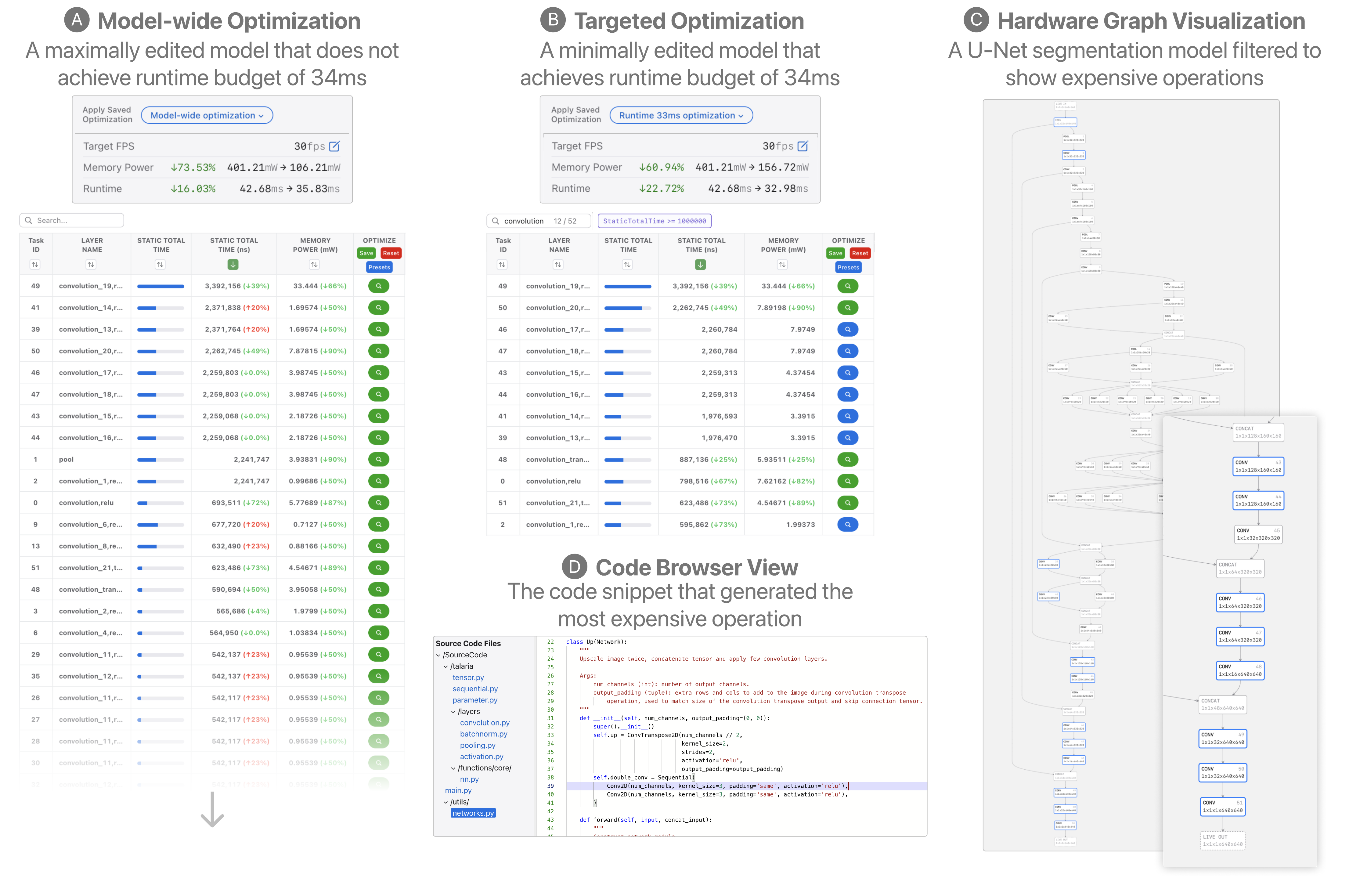}

 \caption{
 An illustrative usage scenario where an ML practitioner \user{} must achieve a runtime budget of 34ms on a U-Net segmentation model.
 With \system{}, she (A) quickly tests a model-wide optimization baseline (using the quantization compression technique, but does not meet budget.
 Instead, she (B) filters the hardware operations to find bottleneck nodes, applies targeted quantization optimization, which meets the budget.
 (C) The \graphview{} highlights the most computationally expensive operations from the earlier filter, and the (D) Code Browser view shows which code snippet generated them.
 }
 \label{fig:usage-scenario}
\end{figure*}

\paragraph{Scenario setup: How to speed up inference of an image segmentation model?}
\user{} is an ML engineer on a product team developing a model that will power a new feature on a mobile device.
The task is image segmentation, and the team decides to use a lightweight U-net architecture~\cite{ronneberger2015u}.
\user{} has been iterating on this model to get the best accuracy possible.
To ship this model on-device, its inference runtime must be within budget to ensure a good user experience.
To start, \user{} loads the model into \system{} to benchmark its current runtime. 
In the system header, she reads off the top-level metrics for the model: ``Memory Power: 401.21mW'' and ``Runtime: 42.68ms.''
The allowed runtime budget for this model is 34ms, so she needs to reduce the runtime by about $20\%$.

\paragraph{Visualizing model architecture on hardware.}
\user{} first familiarizes herself with \system{}, including the two main views: the \tableview{} and \graphview{}.
She sees 51 rows in the \tableview{}, corresponding to 51 model operations running on the hardware.
She first wants to get a sense of how these operations are organized, so in \graphview{} she zooms and pans around the model to inspect the structure generated by the hardware compiler.
She sees the U-Net architecture running on hardware represents her expectations: the input and output share the same size, and the two ``sides of the U'' (called the contracting and expansive paths~\cite{ronneberger2015u}) are seen from the graph connections running from subsequent convolutional layers from the beginning operations to the final operations.

\paragraph{Quick test: Applying model-wide optimizations.}
When analyzing a new model, a common baseline is to try model-wide optimization: optimizing every model operation with the same compression technique.
\user{} wants to see if this quick test satisfies her runtime budget.
She clicks the model-wide optimize button and sees multiple compression options supported by \system{}, including quantization, pruning, and palettization.
\user{} is mainly interested in quantization, so she chooses to cast all input, output, and kernel formats from \texttt{fp16} to \texttt{int8}.
The resulting model (\Cref{fig:usage-scenario}A) reports top-level metrics of reducing memory power by 73.53\% (401.21mW → 106.21mW) and runtime by 16.03\% (42.68ms → 35.83ms).
Note that there is no guarantee that optimizations always make performance better, \eg the overhead of optimization could be larger than the savings.
In this example, the runtime of some operations (colored red in the \tableview{} of \Cref{fig:usage-scenario}A) are increased.
Although this is a big performance improvement, it does not achieve the runtime budget of 34ms. 
Before trying another optimization, \user{} clicks the ``Save'' button and provides a name ``Model-wide optimization,'' to keep a checkpoint of her work.

\paragraph{Analyzing model statistics and finding bottleneck operations.}
Before trying a targeted optimization, \user{} needs a deeper understanding of the model performance.
To inspect model statistics, she reads the \tableview{} to examine existing operations and their runtime distribution.
Scrolling through the tasks and reading down the ``Layer Name'' column, she sees the model is mainly composed of convolution and pooling operations.
From model-wide optimization, she finds quantizing pooling layers does not reduce runtime, so she enters ``convolution'' in search box to focus on these operations.
Since the \graphview{} and \tableview{} are interactively synced, now the \graphview{} highlights the convolution operations with a blue border.
She then sorts the convolution operations by their runtime to reveal the runtime distribution across the model.
From the \tableview{}'s ``Static Total Time'' column, she finds twelve operations take up a majority of the total runtime.
She then applies a filter to remove the operations that are less than 1ms.
Once again, the \graphview{} updates to highlight the convolution nodes that satisfy the filter (\Cref{fig:usage-scenario}C).
These bottleneck operations form the candidate set that \user{} wishes to optimize.

\paragraph{Combining geometric and analytic model knowledge.}
Using the ``Color by Hardware Stats'' feature, \user{} visualizes model architecture and runtime together in \graphview{}.
This feature colors each node a shade of blue (darker means longer runtime).
She confirms that the darker nodes are the operations she has filtered in the \tableview{}, and makes the observation that they appear at the beginning and end of the model. 
This is a fast and powerful way to confirm and visually find model bottlenecks.

\paragraph{Applying targeted model optimizations.}
\user{} now has her candidate set of operations for a targeted optimization.
She clicks the optimize button for the most computationally expensive operation and sees a list of combinations of compression techniques.
\user{} starts with quantizing this operation by filtering the table with \texttt{int8} for the input, output, and kernel; the result shows 39\% reduction of the runtime and 66\% reduction of the memory power for this single operation.
After selecting this option, \system{} applies the optimization and shows \user{} the improvements in the table row.
The top-level metrics in the system header are also updated to show that the overall memory power is reduced by 43.14\% (401.21mW → 228.12mW) and the runtime is reduced by 17.45\% (42.68ms → 35.23ms)---this is close but still not under the required budget (34ms).
\user{} tries to optimize the next most computationally expensive operation with the same quantization.
\system{} updates the metrics and shows an improved memory power reduction of 60.94\% (401.21mW → 156.72mW) and runtime reduction of 22.72\% (42.68ms → 32.98ms).
While this optimization's memory power reduction is not as strong as the model-wide optimization, her targeted optimization (\Cref{fig:usage-scenario}B) successfully meets her runtime budget.
Note that if an operation is dependent upon other operations, \system{} handles these dependencies and optimizes the corresponding operations.\footnote{For example, in \Cref{fig:usage-scenario} the 0th and 49th operations are connected by a path, therefore quantizing the 49th operation's input to \texttt{int8} will update the 0th operation's output to be \texttt{int8}. Similarly, the 51st operation's input must match to \texttt{int8} due to the 50th operation's quantization.}
Before moving on, \user{} clicks the ``Save'' button and names the analysis ``Runtime 33ms optimization.''

\paragraph{Sharing optimized models with others and evaluating on hardware.}
With her targeted optimization and model-wide baseline analyses completed, \user{} wants to share them with her team.
In \system{}, she clicks the share button to add emails of team members, who will see this model in their model lists.
\user{} also copies and pastes the \system{} URL into her team's chat, so others can directly access the model.
Now, other team members can inspect the analysis checkpoints \user{} made, fork and create their own optimizations, and share back with her.
While her team inspects the results, \user{} prepares her code to make the necessary modifications to apply the optimizations.
To locate the code to modify, she clicks on each optimized operation, and then clicks the Code Tracking tab, which highlights the code snippet from the Python source code that generated this hardware operation.
For better context, \user{} clicks on the filename of the snippet to see its location in the codebase (\Cref{fig:usage-scenario}D).
With her code updated, she now can run and evaluate the optimized model on hardware: she finds the actual runtime was reduced to 33.35\%, only around a 1\% difference from the predictions made by \system{}.
\system{} allowed \user{} to understand and experiment, in real-time, with optimizations for her segmentation model, instead of blindly applying compression techniques and waiting longer for hardware benchmarking.

\section{Evaluation: Log Analytics, Usability Survey, and Qualitative Interview}
\label{sec:evaluation}

We deployed \system{} within our organization and over time gained users as multiple teams found it valuable to their work.
We described the system as a new, interactive approach to help ML practitioners evaluate and optimize their model inference efficiency.
Here, we report on three different evaluations (\textbf{E1--E3)}:

\begin{itemize}
    \item[\textbf{E1.}] A \textbf{log analysis} (\Cref{subsec:eval-log}) to track the growth of users and models in \system{} over time.
    \item[\textbf{E2.}] A \textbf{usability survey} (\Cref{subsec:eval-survey}) to determine the most and least useful features to users.
    \item[\textbf{E3.}] A \textbf{qualitative interview} (\Cref{subsec:eval-interview}) with the most active users to learn about their experience using the system for over time and their suggested improvements to help them create efficient ML models.
\end{itemize}

\paragraph{Timeline}
The implementation of \system{} started in the Summer of 2021, with the first version completed in the Fall of 2021.
We have been actively developing the tool since then, including adding features, providing maintenance, and talking with practitioners over 2 years.
The log analysis data was captured from the Fall of 2021 to the Fall of 2023.
The usability survey was sent in the Spring of 2023.
Similarly, for the qualitative interview, we spoke with the power users of \system{} in the Spring of 2023.

\paragraph{Protocol}
Our study includes three evaluations, all of which had their protocols approved by an internal IRB.
Recruitment strategies for each evaluation are described separately in their own section. 
No compensation was given, as all participants were salaried employees of our organization.
However, many participants were interested in learning about our results.
At the end of the study, we briefed participants and their teams on our results.

\subsection{Log Analytics}
\label{subsec:eval-log}
In this first evaluation, we analyze the backend logs of \system{} as one angle to inspect its usage and broader adoption over time.
Inspecting user logs in aggregate gives us insight into the tool's adoption, performance, and user behavior patterns, which can lead to opportunities for future improvements.
In our evaluation, we focus on inspecting cumulative quantities, such as the number of users logged and the number of models submitted.
A deeper analysis, such as which interactions each user takes on specific UI elements, is out of scope for this work.
To protect user privacy, all names have been scrubbed from the data.

\begin{figure}[!b]
 \centering
 
 \includegraphics[width=\columnwidth,alt={Two line charts that show the cumulative total users and models of \system{} over its develop. The user line chart rises steadily for 1.5 years from late 2021 to mid 2023, with two sudden increases towards the middle of 2023. The model line chart rises slowly for 2021, but steadily increases until mid 2023.}]{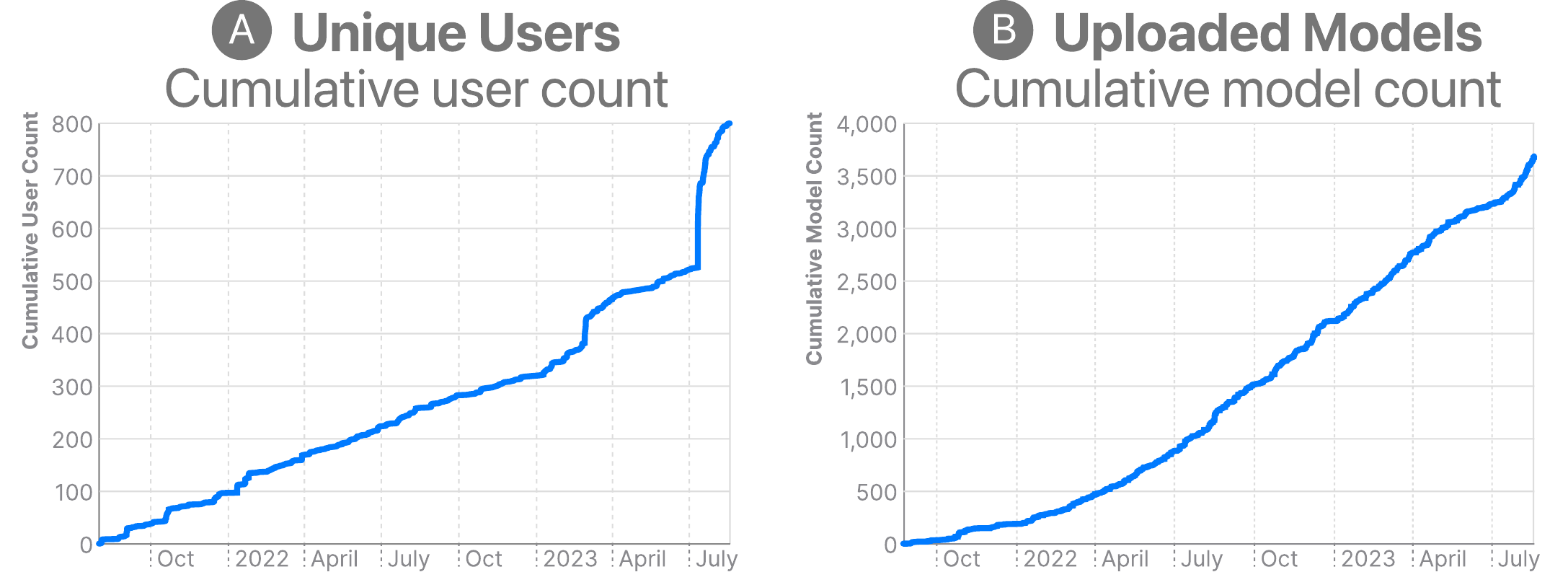}

 \caption{
 The cumulative number of (A) unique \system{} users (800 total) and (B) unique models submitted to \system{} over time (3,600+ submitted).
 }
 \label{fig:log}
\end{figure}

After filtering out the developers of the system and models used for testing, we count 800 unique users, 161 of which have submitted at least one model (20\%).
This means one-fifth of users submit a model, whereas others view a model shared to them by a collaborator. 
Observing the cumulative number of users over time is shown in \Cref{fig:log}A.
Similarly, we can inspect the cumulative number of models that have been submitted.
Over the same time frame, there have been 3,600+ models submitted, as shown in \Cref{fig:log}B.

In both charts in \Cref{fig:log}, we see an interesting pattern: there are multiple large upticks in usage at a single time.
In the users chart in \Cref{fig:log}A, this suggests that an entire team discovered \system{} by viewing a model that was shared with them, or a teammate was demonstrating the tool and had colleagues simultaneously log in to try it organically.
Note that the largest, most recent spike happened when some models were demoed and shared to wider audiences for educational purposes.
In the model chart in \Cref{fig:log}B, upticks suggest that a developer submitted multiple models at once, perhaps testing different hyperparameters or architectures.
These usage patterns are useful vectors for understanding how ML practitioners use \system{}, and are discussion points we follow up on below.

\subsection{User Survey on Feature Usability}
\label{subsec:eval-survey}

In our second evaluation, to understand the usability of \system{}, we surveyed users to rate the usefulness of different system features.
The survey first asked for basic information about a participant's job title, role, and duration / frequency using the system.
The remaining questions asked participants to rate 20 different \system{} features, grouped into the categories described in \Cref{sec:system}.
We piloted the survey with three practitioners to ensure it took less than 5 minutes to complete. 
For recruitment, we sent the survey to email and chat groups specifically related to the tool's development and user base.
In total we received \NumSurveyParticipants responses.

\begin{figure}[tb]
 \centering

 \includegraphics[width=\columnwidth,alt={Three bar charts showing the metadata of the usability survey participants. The first chart shows that a majority of the participants are ML engineers, with a handful of research scientists, hardware engineers, and software engineers in decreasing order. The second chart shows that majority of participants have 5-8 and 9-12 years of experience, followed by 1-4 years, and only one person with 13+ years of experience. The third chart shows most participants use \system{} multiple days per week or weekly, with fewer using it monthly and only a couple using it as needed.}]{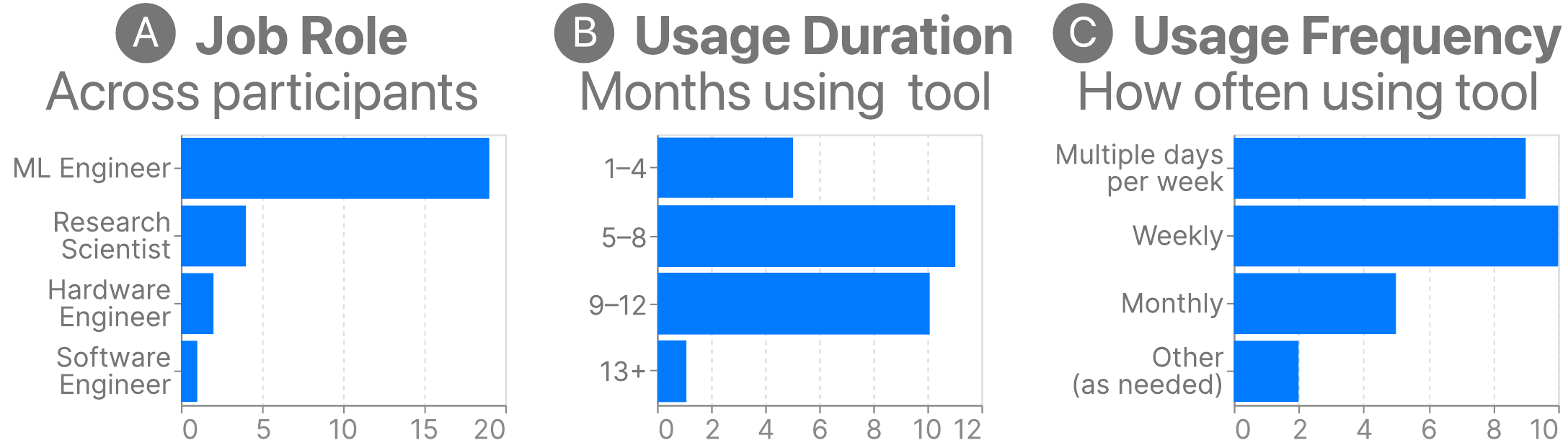}

 \caption{
 A summary of the usability survey participants, including their (A) job role, (B) how long they have used \system{}, and (C) how often they use \system{}.
 }
  \label{fig:survey-participants}
\end{figure}

Our participants, summarized in \Cref{fig:survey-participants}, include multiple types of ML practitioners (\Cref{fig:survey-participants}A), including research scientists, ML engineers, and hardware engineers.
They also span a wide breadth of application domains, such as ML prototyping, model training, model evaluation, hardware, and compiler design.
When asked how long they have used \system{} (\Cref{fig:survey-participants}B), responses ranged from 1 to 18 months.
During that time, when asked how often they use \system{} (\Cref{fig:survey-participants}C), responses showed most practitioners use \system{} multiple times a week or weekly, which is strong evidence that the system has been impactful to their work.

\begin{figure}[t]
 \centering

 \includegraphics[width=\columnwidth,alt={A grouped bar chart colored by responses to the usability survey. Almost all features in the \tableview{}, \graphview{}, and Interactive Model Optimization categories are rated very useful.  Collaborative Optimization and Complementary Visualizations were rated very or somewhat useable, but only 75\% of participants had used those features. Lastly, the Source Code Mapping features had only been used by roughly 50\% of participants, although of those they said it was very useful.}]{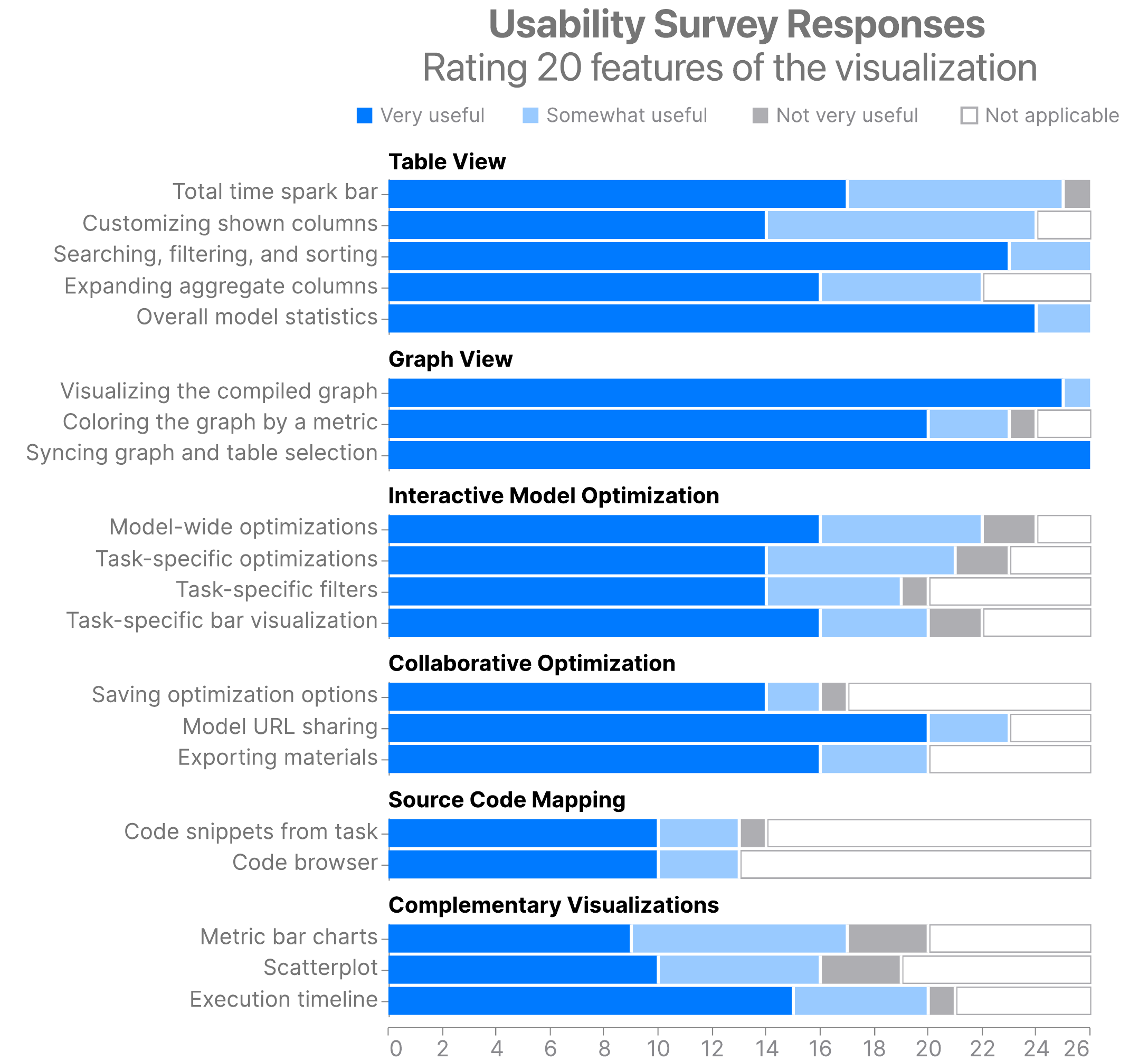}

 \caption{
 The responses to the usability survey grouped by feature.
 Participants rated 20 different features of the system.
 }
 \label{fig:survey}
\end{figure}

Inspecting the responses to the study in \Cref{fig:survey} reveals a number of patterns.
First, in general it is encouraging to see a majority of responses are positive across all feature categories.
Standout features that are the most useful to practitioners include the \tableview{}, \graphview{}, and interactive optimization options.
While the reception to various features within the \tableview{} are high, of the two main views it is surprising how strong the positive response is for the \graphview{}.
This shows the power of visualization: while many optimization tasks can be solved with the \tableview{} (\eg sorting tasks by a particular metric to find the most computationally expensive tasks), viewing a model statistics geometrically by encoding them in the graph provides invaluable context.
It is also encouraging that the complementary visualizations are rated highly useful, despite their conventional design and utility.

If we consider the features that were least useful or not applicable to users, the collaboration and and source code mapping categories stand out.
While both of them have half or more of their responses being very useful, these two categories are the least used or known.
We suspect that not all \system{} users are collaborating within a larger team, and some may use the tool individually.
It also could be the case that a user accomplishes everything they needed within \system{}, and does not need to export any other materials.
The source code mapping features having more not applicable responses is also insightful.
One hypothesis here is that of the two types of optimizations, applying model-wide optimization does not require specific code edits, since the optimization simply applies to every operation; therefore a user does not need this feature.
Another hypothesis is that the discoverability of these features could be improved, since the results show these features are useful or not applicable, only 1 of \NumSurveyParticipants response says they are not useful.

\subsection{Qualitative Feedback from Power Users}
\label{subsec:eval-interview}

In our final evaluation, we gathered feedback during several 30-minute semi-structured interviews~\cite{boyce2006conducting,knott2022interviews} with \system{}'s most active users, \ie power users, to understand their experience of visualizing and optimizing their own models.
We chose a semi-structured format to ensure participants spoke to each question we prepared, with the flexibility to freely speak to their specific work and express any alternative viewpoints or opinions they may hold~\cite{knott2022interviews}.
This method is well-suited to gather firsthand and personal knowledge of efficient ML work that was not captured or anticipated in our previous evaluations~\cite{boyce2006conducting}.

\system{} power users were found by computing the total number of models submitted by each unique user and sorting to find the ones who have submitted the most models. 
We interviewed \NumInterviewParticipants users, including research scientists, ML engineers, and hardware engineers.
A summary of the participants can be found in \Cref{tab:interview-participants}.
These users have interacted with \system{} the most and are already proficient using its features.
We asked specific questions about their user experience, including questions to make them reflect on their own work.
We also asked open-ended questions to learn about future improvements that could help them better optimize their models.
For all interviews, one author led the questioning, while another took notes.
With participant's approval, we recorded conversations to refer back to during analysis.

The interview questions were structured around the challenges that practitioners face with efficient ML (\Cref{sec:design-challenges}) and tasks we identified that tooling should support (\Cref{sec:tasks}).
From the interview data, we conducted a thematic analysis method to group common workflows, user behavior, and best practices of model optimization into categories~\cite{gibbs2007thematic}.
Each participant's data and transcripts were independently reviewed and manually coded using inductive coding~\cite{thomas2003general}.

\subsubsection{Analytically and Visually Optimizing Models}

It was exciting to learn that practitioners had their own preferences for the views they used in their analyses.
Between the two main views (\tableview{} and \graphview{}), their preference was nearly split: after uploading a new model, P2, P4, and P6 looked at the \tableview{} first, whereas P1, P3, P5, and P7 considered the \graphview{} first.
Despite this first reaction, nearly all participants mentioned that they relied on two views together for analysis (\textbf{T1}).
P4 stated it plainly: \textit{``Both the numbers and graph are equally important.''}
Participants told us that selecting a task in the \tableview{} and simultaneously highlighting it in the \graphview{} (and vice versa) was transformative to their work.
Of all the features in \system{}, P2 said this interactive selection between the views was their favorite.

One unexpected task supported by the \graphview{} was that practitioners used the graph to verify architecture questions they had when building a model.
This is likely a potential reason that the \graphview{} was rated so highly in the usability survey (\Cref{subsec:eval-survey}).
For example, P3 said that they use the graph to confirm their understanding of an architecture change, and are then eager to see how it compiles to hardware.
P2 said they view the graph as a \textit{``quick check.''}
This model verification task is interesting, as it emphasizes the unique consideration of hardware details that conventional ML does not usually need to work with.
To measure on-device metrics such as power, latency, and memory usage, practitioners need to know how their models will decompose into individual operations on hardware.
Visualization greatly helps in this task by allowing practitioners to visually inspect the topology of their model graphs and to encode different metrics on top of the graph.

\begin{table}
\centering
\caption{
A summary of the participants interviewed for the qualitative interview evaluation, including their roles, primary types of ML application, and years of experience.
}
\label{tab:interview-participants}
\begin{tabular}{clll}
\textbf{ID} & \textbf{Role} & \textbf{ML Application} & \textbf{Exp.}\\
\midrule
P1 & Research Scientist & Research \& Optimization & 6 yrs\\
P2 & Hardware Engineer &  Deployment \& Optimization & 5\\
P3 & ML Engineer & Training \& Optimization & 6\\
P4 & ML Engineer & Training \& Optimization & 6\\
P5 & Research Scientist & Research \& Optimization & 4\\
P6 & ML Engineer & Optimization & 7\\
P7 & ML Engineer & Training \& Optimization & 7\\
\end{tabular}
\end{table}

\begin{quote}
    \textit{``I use \system{} to sketch out the topology of a model; it is a nice tool to visualize a model as well as looking at the power and perf.''} --- P7

\end{quote}

\subsubsection{Discovering Computational Bottlenecks}

We next asked about \system{}'s ability to find computational bottlenecks (\textbf{T2}), or what P2 referred to as \textit{``top offenders''} and P7 referred to as \textit{``hot spots''} (\ie tasks that have the most latency, memory, or power consumption). 
A major goal of the \system{} design was to allow practitioners to find model bottlenecks quickly, either from low-level statistics, the model graph, or other visualizations. 
It was unsurprising then that all participants said this was one of their primary reasons to use the tool, and that \system{} did it well.
We dig into the bottleneck finding process by asking if practitioners had ever uploaded a model and been surprised by a bottleneck.
P1 said this \textit{``happens often,''} and P2 said this \textit{``happens all the time.''}
More specifically, P3, P4, and P5 said that they have all uploaded models and found additional hardware tasks that were not supposed to be there.
For example, when applying a targeted quantization to a subset of hardware tasks, practitioners found redundant data type conversions between the input and output of various hardware tasks.
With \system{}, they could find these bottlenecks and fix them faster than before.
\begin{quote}
\textit{``The nice thing about \system{} is that it tells you stuff that you might not be expecting, but it also gives you a way to see why that was happening.''} --- P2
\end{quote}

\subsubsection{Faster Optimization Experimentation}

Beyond visualizing model statistics and finding computational bottlenecks, we investigated how the power users engaged with the interactive optimization features (\textbf{T3}).
Use cases here varied by practitioner needs.
For example, P6 heavily uses the model-wide optimization.
P6 works with and consults for multiple model development teams, so whenever they receive a new model, they need the fastest way to test the maximal savings to quickly share back to the teams, which can be achieved by optimizing an entire model with a particular compression technique.
The other six participants more often use the targeted optimization features.
Based on their applications, participants preferred different compression techniques (\eg quantizing inputs and outputs only, quantizing kernels, or pruning weights).
P3 said they appreciate that \system{} \textit{``clearly shows me what options I have for each layer.''}

\begin{quote}
\textit{``\system{} is nice because I can try a couple of optimization options quickly, and it can tell me at a finer level what's going on.''} --- P7
\end{quote}

One unique workflow worth highlighting was from P4, where they said they prefer to do targeted optimization because they do not want to change every layer, which is more likely to cause accuracy loss.
P4 instead works backwards, by applying model-wide optimization first and then removes optimizations to the sensitive layers that need to be preserved.
We noted this approach to inform future users that they can optimize the full model but also selectively remove tasks that need full precision.

\subsubsection{Optimizing Models within Teams}

We also asked about the practitioners experience using \system{} in a collaborative setting (\textbf{T4}). 
From the interviews, it was clear that sharing is heavily used, but we also wanted to better understand the model receivers: are they modeling engineers, hardware experts, or broader stakeholders?
When sharing \system{} URLs within their own team, P2 said they will iterate on models individually and then share the best model as final proof of their work.
P7 has a similar workflow, where when they receive a new model, they upload it to \system{}, then send back a \system{} URL to their collaborators, saying: \textit{``This is what you originally had, and here's what I got it down too.''}
P3 and P5 said they will share multiple URLs (different versions of a model) to their teams for comparison.
P4 and P6 said that compared to only reporting top-level metrics, it can be more valuable to share \system{} URLs in case a stakeholder wants to go deeper.

Lastly, P1 recounted a scenario where they were consulting for reducing model latency.
They found themselves in-between a modeling team and a hardware team, and regularly shared \system{} URLs to both teams to explain changes and potential savings.
P1, an efficient ML expert, explained that they regularly consult on projects that need to hit tight budgets to produce the best user experience.
While they gladly share their expertise, this approach is not scalable, especially as the number of projects grow.
They were excited to see interactive tools, such as \system{}, help others without this expertise optimize their own models.

\begin{quote}
\textit{``Since some people have [efficient ML] tribal knowledge, [...] self-service is definitely the future.''} --- P6
\end{quote}

\subsubsection{Closing the Loop: Applying Optimizations}

Lastly, we report on practitioners taking their optimization analysis and applying it back to their codebase (\textbf{T5}).
Recall in the usability survey this feature category was the least used (\Cref{fig:survey}).
This result is also reflected in our interviews, where practitioners did not have as many examples to describe.
Our original intent was that practitioners have an actionable next step after using \system{}.
Our novel contribution here is attributing individual hardware operations back to source code.
However, practitioners explained that applying optimizations to code is only one iteration they might do. 
Other iterations a practitioner might do may be trying a different architecture, updating the model compiler, or exporting statistics to run their own additional analysis outside of \system{}.
We believe there is opportunity here to further improve the ML developer experience, however, what is most important is that our users did not get stuck when using \system{}, and that the system gave them something actionable to do next, even if it was not within the system itself.
\begin{quote}
    \textit{``Ultimately \system{} helps in creating models that run faster, while being more friendly to the developer.''} --- P6
\end{quote}

\section{Discussion: Limitations and Future Work for Optimization Visualization}
\label{sec:future-work}

\begin{figure*}
 \centering
 
 \includegraphics[width=\textwidth,alt={Two screenshots of the new \diffview{} view in \system{}. The first screenshot shows the source code for a segmentation model, and the modified source code for a new model with two additional lines of code representing additional layers in the neural network. The second screenshot shows the main \system{} UI but split to show two models. This includes two tables and two computational graphs, where rows of the table and nodes of the graph are colored green for the new operations spawned from the modified code.}]{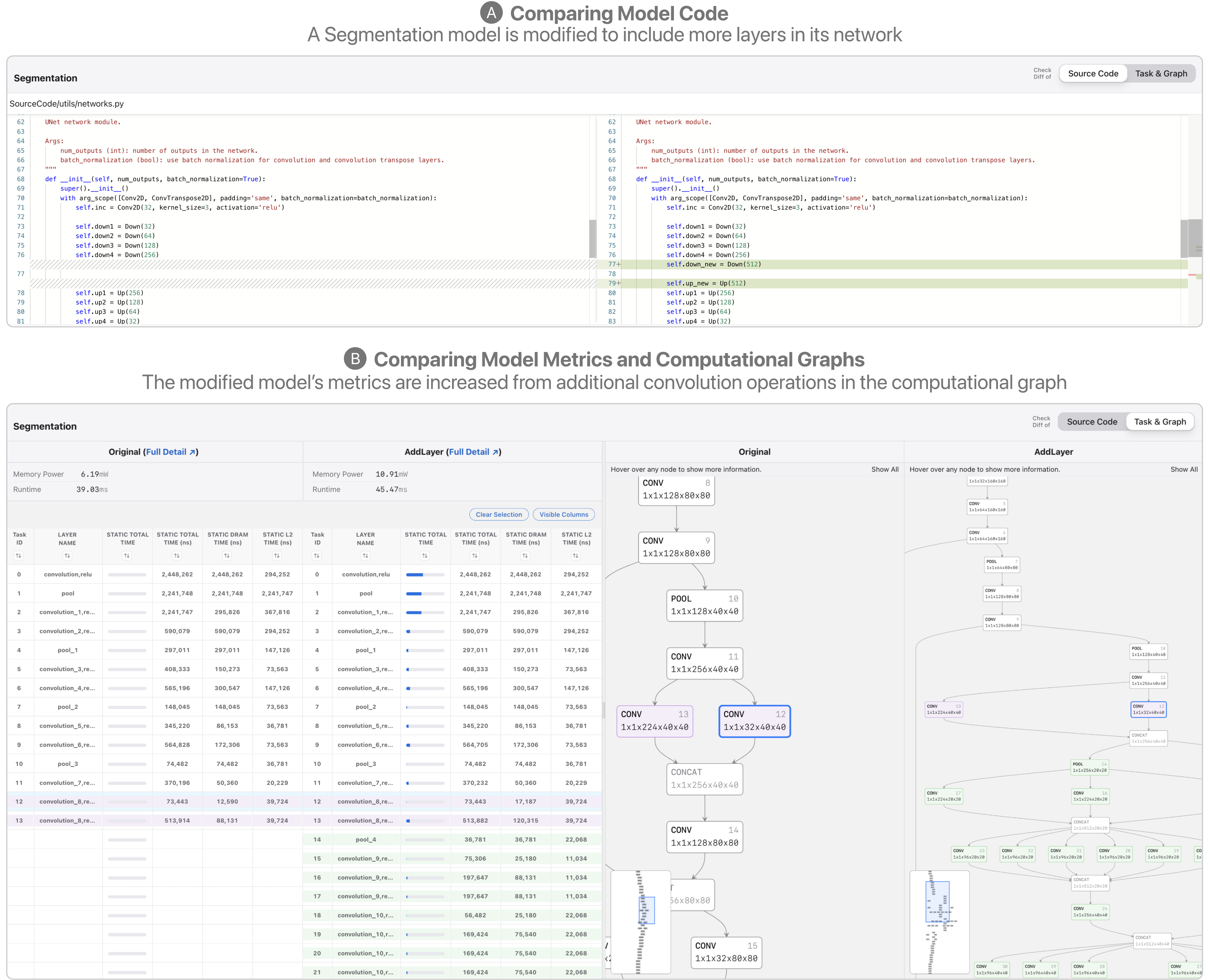}

 \caption{
 The prototype model \diffview{} added after observing practitioners from our evaluation comparing multiple models in \system{}.
 In this example, (A) a ``Segmentation'' model's code is modified to include additional layers in its network.
 (B) The new view shows both the original model and the modified model's hardware statistics and computational graphs, highlighting new operations in green.
 This new model adds multiple convolutional layers to the graph, which increases the memory power from 6.19mW to 10.91mW, and the runtime from 39.03ms to 45.47ms.
 }
 \label{fig:diff}
\end{figure*}

\subsection{Model Comparison}
From our log analysis in \Cref{subsec:eval-log}, we observed a particular user behavior: ML practitioners may submit multiple versions of a model at once for comparison.
A limitation of \system{} is that it only visualizes one model at a time; however, ML development is highly iterative and experimental~\cite{patel2008investigating,amershi2019software}, requiring practitioners to compare model statistics, architectures, and hyperparameters.
Efficient ML work adds another piece to this puzzle, as practitioners also need to consider trade-offs between hardware metrics, such as model size, power, and latency.
From our qualitative study in \Cref{subsec:eval-interview}, users want to compare models across multiple facets.
Example comparisons include comparing an optimized model to a non-optimized model, comparing different compression strategies, or comparing models with different architectures altogether.
This introduces new challenges: how should models be compared, \eg against a common baseline or against one another?
How do we effectively visualize relevant differences between models?
What if a user wants to compare more than two models?

Since this observed workflow was so important and prevalent, after our study analysis concluded we implemented a new prototype view into \system{} called the model \diffview{}.
While this view does not fully support arbitrary and flexible model comparison, it does help practitioners with the common task of comparing two models, their hardware metrics, and their computational graphs, against one another.
As seen in \Cref{fig:diff}A, the code for a model on the left is modified, and a new model with additional layers is created on the right.
With both models loaded into \system{}, the \diffview{} now divides the main interface into four sections: two tables on left and two computational graphs on the right, mimicking the \tableview{} and \graphview{} for inspecting a single model.
In the updated \tableview{}, \Cref{fig:diff}B shows new layers that are not present in the original model highlighted in green, and layers that were removed highlighted in red (none present in this example).
Similarly, the updated \graphview{} shows both computational graphs, with new hardware operations colored green.
With this new view, practitioners can see what impacts different model architectures have on their top-level metrics, and where modified hardware operations are located in the model's computational graph.

This is an early exploration into model comparison for ML optimization.
It is important to note that model comparison visualization is not a new topic and has been explored in other tools~\cite{xuan2022vac, das2020legion, kahng2016visual}.
However, given the size and complexity of modern ML models, improved visualizations for model comparison is worth revisiting, especially for the new challenges and constraints brought with efficient ML.

\subsection{Automatic Code Editing and Interactive Model Playgrounds}
\system{} allows users to test various optimization options and inspect their impact on inference efficiency.
However, right now a practitioner must still manually apply those optimizations in their code.
\system{}, or future tools for model compression, could automatically apply the specified optimizations in code (possibly using large language models pretrained for coding tasks~\cite{copilot,codex,gpt4}), recompile them to the targeted hardware, and visualize the results.
Drawing inspiration from fluid end-user programming tools that sync code and GUI states~\cite{kery2020mage}, we propose an interactive playground where users upload their initial model definition code, iteratively apply optimizations, recompile their models, and finally use the optimized model code for retraining.

Lastly, given that \system{} contains both a model's code and available optimization options, there is opportunity to automatically suggest recommended compression techniques to try first.
Recommending compression techniques may sound appropriate for an automated optimization algorithm.
However, fully automating model optimization is not yet possible, due to how many considerations must be made both about the model and the design of the user experience the model will enable~\cite{hohman2024compression}.
Nevertheless, future tools could enable mixed-initiative interaction and guided experimentation, where \system{} could have the power to recommend optimization options in the interface to a user and make changes to a model's source code.
These feature additions could save practitioners a significant amount of time, providing more opportunities to iterate on their models.

\subsection{Including Model Behavioral Metrics}
\system{}'s focuses on improving the inference efficiency of ML models running on-device.
While it is possible to apply maximal compression to extremely optimize model efficiency and hardware metrics (\eg model size, latency, and power), it may negatively impact the model's behavioral metrics (\eg accuracy, precision, recall).
The holistic goal of building efficient models is to find a balance between inference efficiency and an acceptable accuracy regression.
One limitation of \system{} is that it currently does not take into account model behavioral metrics such as accuracy, and instead focuses specifically on the new and novel challenges brought with efficient ML work.
Today with \system{}, a practitioner could quickly apply maximal optimization and minimal optimization to a model, then retrain them with these optimization configurations to check how the accuracy or other behavioral metrics changed.
However, there is great opportunity to combine \system{} more deeply with model evaluation tools that visualize behavioral metrics across different subgroups of data (\eg to catch potential fairness or accessibility concerns).

Certain technical challenges will need to be addressed to do these evaluations in real-time for interactivity, since considering behavioral metrics requires a forward pass of one's testing data through the model to compute predictions.
Depending on the size of the test set, or the size of the model, this may take on the order of minutes to hours.
Perhaps applying bootstrap sampling methods to create ``efficient ML test sets'' that a model could predict over in seconds would allow future tools to test certain model optimizations and get both behavioral and hardware metrics in real-time. 
This potential combination would allow ML practitioners to easily see the impact that compression methods have on behavioral metrics and inference efficiency simultaneously.

\subsection{Collaborative Model Optimization}

While \system{} enables practitioners to save optimization experiments and share them with others, its collaborative features are lightweight compared to other feature sets.
\Cref{subsec:eval-survey} shows that the existing features are rated highly useful, but this is only a first step in the direction of collaborative, efficient ML.
Collaboration in data science is not a new topic.
Popular programming tools have embraced collaborative features, such as Juypter~\cite{kluyver2016jupyter}, Google Colab~\cite{bisong2019google}, and VSCode~\cite{microsoft2023visual}, and previous work has profiled how data scientist work collaboratively, both in interpersonal relationships and with tools~\cite{zhang2020data, randles2017using}.
\system{} supports collaborative tooling design highlighted by \citet{zhang2020data} by capturing the end result of an analysis with code and documentation (\eg saving shareable optimization analyses and model metadata), but future extensions could see additional support for tracking a full history of one's analysis~\cite{kery2019towards, head2019managing}.
Historical, collaborative features could help others reproduce an optimization step-by-step to support better reproducibility---a critical challenge due to the iterative, empirical nature of ML work~\cite{patel2008investigating, amershi2019software} that model optimization further complicates with additional dimensions such as compiler versions, hardware targets, and compression techniques.

\subsection{Scaling Visualization Design}

\system{} was built with scalability in mind, particularly for large, modern ML models.
While we have not done an exhaustive scalability test, \system{} has been used for models with thousands of tasks/graph nodes and runs smoothly.
The \tableview{} only renders rows within the browser's viewport, making scrolling, sorting, filtering, and searching in real time possible even for large models.
Zooming and panning on the the \graphview{} is fast, since the graph is rendered on canvas using WebGL and runs at a high refresh rate (\eg 60fps) even with thousands of nodes.

However, we have tested some models that had tens of thousands of hardware operations.
In these models, the \graphview{} was usable, but the bigger challenge in navigating the graph was that it was too large to get an intuitive sense of how the model compiled onto hardware.
A good example of this is visualizing a transformer model, where the thousands of operations could be alternatively represented as a handful of sequential transformer modules.
In this regime of scale, future visualization and interaction design could help, for example, by exploiting repeatable hardware operation types and automatically grouping them into supernodes (similar to~\cite{wongsuphasawat2018visualizing}).
While users can define their own groups in code before submitting models to \system{}, in the future groups could be constructed automatically based on exploiting repeatable hardware operations, either in sequence such as multiple convolution operations, or mined as patterns across a model (\eg a parallel convolution structure that concatenates into a pooling operation).

\subsection{Future Tools for Efficient ML}

The goal of this work was to show evidence of how interactive tooling for ML optimization can be highly productive in practice.
Reflecting on our evaluations, one characteristic that stands out from \system{} compared to previous work is the effort to unify the existing scripts, views, and ad-hoc analyses of practitioner workflows into single system paid off.
\system{} lowers the barrier to efficient ML work and makes optimization estimation easier (\eg clicking a button), helping people inspect the trade-offs between multiple model optimizations.
This holistic view of efficient ML work, combining hardware and software, is a key differentiator between \system{} and existing work.

The design of \system{} was guided by our formative research with expert ML practitioners.
We followed known visualization design patterns~\cite{brehmer2013multi}, such as implementing multi-coordinated views, cross-filtering, and Schneiderman's mantra~\cite{shneiderman1996eyes} for overview + detail and focus + context techniques~\cite{cockburn2009review} for mixed-initiative user interfaces~\cite{horvitz1999principles}.
Despite having rigorous strategies for designing interfaces, we emphasize that tooling in efficient ML is currently underdeveloped and underexplored~\cite{hohman2024compression}.
The few related tools focus on explaining the inner workings of a particular compression algorithm (\Cref{subsec:vis-for-opt}).
While existing work advances our understanding of specific techniques, they may not be generalizable enough for many real-world applications.
Future work on designing tools for efficient ML have abundant opportunity for building on top of rich literature in HCI and visualization to advance the state-of-the-art.

\section{Conclusion}
\label{sec:conclusion}

By focusing on creating on-device and efficient models, we can design new and intelligent ML user experiences.
This direction of research, while growing, is still in its infancy.
More specifically, tooling for creating and optimizing models is underdeveloped.
To help ML practitioners create efficient models, we designed and developed \system{}, an interactive visualization system, alongside ML experts at \location that specialize in developing on-device models.
Our visualization system enables ML practitioners to analyze models across a variety of low-level statistics, interact with a model's computational graph, and experiment with model optimizations on hardware.
We hope our work emphasizes the need and importance of tooling for model optimization, and inspires future work on interactive tooling for creating efficient ML user experiences.

\begin{acks}
The authors thank our colleagues at \location for their energy, support, and guidance over this work.
We especially thank Sam Xu, Matthew Kay Fei Lee, Patrick Dong, and Hojin Kee for their technical expertise.
We also thank those who took time to participant in our system evaluations.

\end{acks}

\balance
\bibliographystyle{ACM-Reference-Format}
\bibliography{main}

\end{document}